\pdfoutput=1
\documentclass[table,xcdraw,smallextended]{svjour3}

\usepackage{etex}
\usepackage{graphicx}
\usepackage{xcolor}
\usepackage[utf8]{inputenc}
\usepackage{listings}  
\usepackage{amsmath}
\usepackage{balance}
\usepackage{tabularx}
\usepackage{booktabs}% for \midrule
\usepackage{mdframed} 
\usepackage{verbatim} % adds environment for commenting out blocks of text & for better verbatim
\usepackage{paralist}
\usepackage{listings}
\usepackage{courier}
\usepackage{xspace}
\usepackage{balance}
\usepackage{algorithm} 
\usepackage{algorithmic}
\usepackage{multirow}
\definecolor{ForestGreen}{RGB}{34,139,34}
\definecolor{lightgray}{RGB}{225,225,225}
\usepackage[]{pdfcomment}
\usepackage{float}
\usepackage{lscape}
\usepackage{rotating}

\usepackage[normalem]{ulem}

\newcolumntype{a}{>{\columncolor{lightgray}}l}
\newcolumntype{b}{>{\columncolor{white}}l}

\newcommand{\rev}[2]{\textcolor{blue}{#2}\GenericWarning{}{LaTeX Warning: Rev#1: #2}}\newcommand\REV\rev

\newcolumntype{s}{>{\hsize=.75\hsize}X}
\newcolumntype{m}{>{\hsize=1.25\hsize}X}
\newcolumntype{L}[1]{>{\raggedright\arraybackslash}p{#1}}
\newcolumntype{C}[1]{>{\centering\arraybackslash}p{#1}}
\newcolumntype{R}[1]{>{\raggedleft\arraybackslash}p{#1}}

\newcommand{\Iampl}{\emph{I-Amplification}\xspace}
\newcommand{\Aampl}{\emph{A-Amplification}\xspace}
\newcommand{\TODO}[1]{\textcolor{red}{#1}\GenericWarning{}{LaTeX Warning: TODO: #1}}\newcommand\todo\TODO
\newcommand{\etal}{\textit{et al.}\xspace}
\newcommand{\ie}{\textit{i.e.}\xspace}
\newcommand{\eg}{\textit{e.g.}\xspace}
\newcommand{\dspot}{DSpot\xspace}
\newcommand{\gh}{GitHub\xspace}
\newcommand{\pitest}{Pitest\xspace}

\newcommand{\ms}{mutation score\xspace}
\newcommand{\ams}{number of killed mutants\xspace}
\newcommand{\junit}{JUnit\xspace}

\newcommand*\rotverticalinv{\rotatebox{90}}

\usepackage{amssymb}% http://ctan.org/pkg/amssymb
\usepackage{pifont}% http://ctan.org/pkg/pifont

\usepackage{listings}
\usepackage{lstlinebgrd}

\newcommand{\bglinediff}[3]{%
        \ifnum\value{lstnumber}>#1
            \ifnum\value{lstnumber}<#2
                \color{#3}
            \fi
        \fi}

\title{Automatic Test Improvement with DSpot:\\ a Study with Ten Mature Open-Source Projects}

\author{Benjamin Danglot \and Oscar Luis Vera-P\'erez \and Benoit Baudry \and Martin Monperrus}
\institute{B. Danglot \at
              Inria Lille - Nord Europe \\
                Parc scientifique de la Haute Borne \\ 
                40, avenue Halley - Bât A - Park Plaza \\
                59650 Villeneuve d'Ascq - France\\
                \email{danglot@inria.fr}
           \and
           O. Vera-P\'erez \at
           Inria Rennes - Bretagne Atlantique \\
           Campus de Beaulieu, 263 Avenue Général Leclerc \\
           35042 Rennes - France \\
           \email{oscar.vera-perez@inria.fr}
           \and
           B. Baudry \at
           KTH Royal Institute of Technology in Stockholm\\
           Brinellvägen 8\\
           114 28 Stockholm - Sweden\\
           \email{baudry@kth.se}
           \and
           M. Monperrus \at
           KTH Royal Institute of Technology in Stockholm\\
           Brinellvägen 8\\
           114 28 Stockholm - Sweden\\
           \email{martin.monperrus@csc.kth.se}
}

\date{March 2018}

\begin{document}

\maketitle

\begin{abstract}

In the literature, there is a rather clear segregation between manually written tests by developers and automatically generated ones.
In this paper, we explore a third solution:
to automatically improve existing test cases written by developers.
We present the concept, design and implementation of a system called \dspot, that takes developer-written test cases as input (\junit tests in Java) and synthesizes improved versions of them as output. Those test improvements are given back to developers as patches or pull requests, that can be directly integrated in the main branch of the test code base.
We have evaluated \dspot in a deep, systematic manner over 40 real-world unit test classes from 10 notable and open-source software projects.
We have amplified all test methods from those 40 unit test classes.
In 26/40 cases, \dspot is able to automatically improve the test under study, by triggering new behaviors and adding new valuable assertions.
Next, for ten projects under consideration, we have proposed a test improvement automatically synthesized by \dspot to the lead developers. 
In total, 13/19 proposed test improvements were accepted by the developers and merged into the main code base. 
This shows that \dspot  is capable of automatically improving unit-tests in real-world, large scale Java software.

\end{abstract}

% ---------------------------------------------------------------------------------------
% INTRODUCTION
% ---------------------------------------------------------------------------------------
\section{Introduction}

% context 
Over the last decade, strong unit testing has become an essential component of any serious software project, whether in industry or academia. The agile development movement has contributed to this cultural change with the global dissemination of test-driven development techniques \cite{beck2003test}.
More recently, the DevOps movement has further strengthened the testing practice with an emphasis on continuous and automated testing \cite{Roche2013Devops}. 

In this paper we study how such modern test suites can benefit from the major results of automatic test generation research.
We explore whether one can automatically improve tests written by humans, an activity that can be called ``automatic test improvement''. 
There are few works in this area: the closest related techniques are those that consider manually written tests as the starting point for an automatic test generation process \cite{Harder03,fraser2012seed,Xuan:2014:TCP:2635868.2635906,Yoo:2012:TDR:2237756.2237758,danglot2017emerging,Xuan:2015:CRV:2786805.2803206}. To this extent, automatic test improvement can be seen as forming a sub-field of test generation.
Automatic test improvement aims at synthesizing modifications of existing test cases, where those modifications are meant to be presented to developers. As such, the modifications must be deemed relevant by the developers themselves (the corollary being that they should not only maximize some criterion). 

For our original study of automatic test improvement, we have developed \dspot, a tool for automatic test improvement in Java.  \dspot  adapts and combines two notable test generation techniques: evolutionary test construction \cite{tonella} and regression oracle generation \cite{Xie2006}. The essential adaptation consists in starting the generation process from the full-fledged abstract syntax trees of manually written test cases. The combination of both techniques is essential so that changes in the setup together are captured by changes in the assertion part of tests. 

Our study considers 10 mature Java open source projects. It focuses  on three points that have little, or never, been assessed. First, we propose 19 test improvements  generated by \dspot to the developers of the considered open source projects. We present them  the improvement in the form of pull requests, and we ask them whether they would like to merge the test improvements in the main repository. In this part of the study, we extensively discuss their feedback, to help the research community understand the nature of good test improvements. This reveals the key role of case studies, as presented by Flyvberg \cite{flyvbjerg2006}, to assess the relevance of our technique for developers.
Second, we perform a quantitative assessment of the improvements of 40 real-world test classes from our set of 10 open-source projects. In particular, we consider the difficult case of improving strong test classes.
Third, we explore the relative contribution of evolutionary test construction and of assertion generation in the improvement process.

Our key results are as follows: first, thirteen \gh pull requests consisting of automatic test improvements have been definitively accepted by the developers; second, an interesting empirical fact is that \dspot has been able to improve a test class with a 99\% initial \ms (\ie a really strong test); and finally, our experiment shows that valuable test improvements can be obtained within minutes.

To sum up, our contributions are:
\begin{itemize}
\item  \dspot, a system that performs automatic test improvement of Java unit tests; 

\item the design and execution of an experiment to assess the relevance of automatically improved tests, based on feedback from the developers of mature projects;

\item a large scale quantitative study of the improvement of 40 real-world test classes taken from 10 mature open-source Java projects.

\item fully open-science code and data: both \dspot\footnote{\url{https://github.com/STAMP-project/dspot/}} and our experimental data are made publicly available for future research\footnote{\url{https://github.com/STAMP-project/dspot-experiments/}}

\end{itemize}

The remainder of this article is as follows.
Section~\ref{sec:dspot} presents the main concepts of automatic test improvement and \dspot.
Section~\ref{sec:protocol} presents the experimental protocol of our study.
Section~\ref{sec:results} analyses our empirical results. 
Section~\ref{sec:threats} discusses the threats to validity.
Section~\ref{sec:related} discusses the related work.
and Section~\ref{sec:conclusion} concludes the article.
Note that a previous version of this paper can be found as Arxiv's working paper~\cite{BaudryARM15}.

%%%%%%%%%%%%%%%%%%%%%%%%%%%%%%%%%%%%%%%%%%%%%%%%%%%%%%%
% CONTRIBUTION
%%%%%%%%%%%%%%%%%%%%%%%%%%%%%%%%%%%%%%%%%%%%%%%%%%%%%%%
\section{Automatic Test Improvement}
\label{sec:dspot}

In this section, we present the concept of automated test improvement, and its realization in the \dspot tool.

\subsection{Goal}

% concept without dspot
The goal of automatic test improvement is to synthesize modifications to existing test cases to increase test quality.
These modifications are meant to be given to developers and committed to the main test code repository.
The quality assessment is driven by a specific test criterion such as branch coverage or \ms.
In this paper, we focus on improving the \ms of an existing test suite but automatic test improvement is more general and it is not bound to the \ms.

\subsection{\dspot}

\dspot is an  automatic test improvement tool for Java unit tests. It is built upon the algorithms of Tonella \cite{tonella} and Xie \cite{TaoXie2006}.

\subsubsection{\dspot inputs}

The input of \dspot consists in a set of existing test cases, manually written by the developers.
As output, \dspot produces variants of the given test cases. These variants are meant to be added to the test suite. By putting together existing test cases and their variants, we aim at strictly improving the overall test suite quality. By construction, the enhanced test suite is at least as good, or better than the original one w.r.t. the considered criterion.

\begin{figure}[!ht]
    \centering\fbox{\includegraphics[width=0.98\textwidth, trim=2cm 17.02cm 4.09cm 10.46cm, clip]{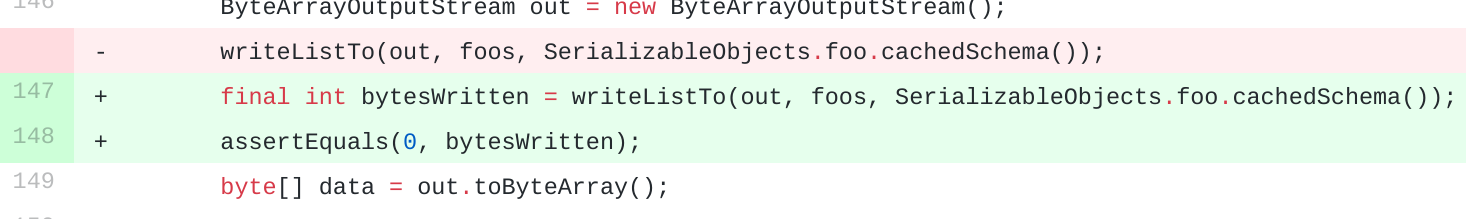}}
    \caption{Example of what \dspot produces: a diff to improve an existing test case.}
    \label{fig:diff-protostuff}
\end{figure}

Concretely, \dspot synthesizes suggestions in the form of diffs that are proposed to the developer: \autoref{fig:diff-protostuff} shows such a test improvement.  

\dspot is an automatic test improvement system because it only modifies existing test cases. As such, all test improvements, by construction, are modifications to existing test cases.
\dspot's novelty is twofold: 1) first, it combines those algorithms in a way that scales to modern, large Java software 2) second, it makes no assumption on the tests to be improved, and works with any arbitrary \junit test.

\subsubsection{\dspot's Workflow}

The main workflow of \dspot{} is composed of 2 main phases:
1) the transformation of the test code to create new test inputs inspired by Tonella's technique, we call this ``input space exploration''; this phase consists in changing new test values and objects and adding new method calls, the underlying details will be explained in details in \autoref{subsec:input-space-exploration}.
2) the addition of new assertions per Xie's technique~\cite{TaoXie2006}, we call this phase ``assertion improvement''. The behavior of the system under test is considered as the oracle of the assertion, see \autoref{subsec:new-assertions}.
In \dspot, the combination of both techniques, \ie the combination of input space exploration and assertion improvement is called ``test amplification''.

% use of mutation score
\dspot keeps the modifications that add the most value for the developers.
To do so, \dspot uses the \ms as a proxy to the developers’ assessed value of quality.
In essence, developers value changes in test code if they enable them to catch new bugs, that is if the improved test better specifies a piece of code.
This is also reflected in the \ms: if the \ms increases, it means that a code element, say a statement, is better specified than before. 
In other words, \dspot uses the \ms to steer the test case improvement, following the original conclusions of DeMillo \etal who observed that mutants provide hints to generate test data \cite{demillo1978hints}.
To sum up, \dspot aims at producing better tests that have a higher potential to catch bugs.

% ---------------------------------------------------------------------------------------
% definitions
% ---------------------------------------------------------------------------------------

\subsection{Definitions}

We first define the core terminology of \dspot in the context of object-oriented Java programs. 

\textbf{Test suite} is a set of test classes.

\textbf{Test class} is a class that contains test methods. A test class is neither deployed nor executed in production.

\textbf{Test method} or \textbf{test case} is a method that sets up the system under test into a specific state and checks that the actual state at the end of the method execution is the expected state.

\textbf{Unit test} is a test method that specifies a targeted behavior of a program. Unit tests are usually independent of each other and execute a small portion of the code, i.e. a single unit or a single component of the whole system.

\begin{lstlisting}[caption={An example of an object-oriented test case  (inspired from Apache Commons Collections)},label=lst:archetype,float,language=java,numbers=left] 
testIterationOrder() {
  // contract: the iteration order is the same as the insertion order
  
  TreeList tl=new TreeList(); (*@\label{input-begin}@*)
  tl.add(1);
  tl.add(2); (*@\label{input-end}@*)

  ListIterator it = tl.listIterator();(*@\label{test}@*)

  // assertions
  assertEquals(1, it.next().intValue());(*@\label{assertion-1}@*)
  assertEquals(2, it.next().intValue());(*@\label{assertion-2}@*)
}
\end{lstlisting}

\subsubsection{Modern Test Cases}
\label{subsec:test-case-explanation}

\dspot improves the test cases of modern Java programs, which are typically composed of two parts: input setup and assertions. 
The input setup part is responsible for driving the program into a specific state.
For instance, one creates objects and invokes methods on them to produce a specific state.

The assertion part is responsible for assessing that the actual behavior of the program corresponds to the expected behavior, the latter being called the oracle.
To do so, the assertion uses the state of the program, \ie all the observable values of the program, and compare it to expected values written by developers.
If the actual observed values of the program state and the oracle are different (or if an exception is thrown), the test fails and the program is considered as incorrect.

\autoref{lst:archetype} illustrates an archetypal example of such a test case: 
first, from line \autoref{input-begin} to line \autoref{input-end}, the test input is created through a sequence of object creations and method calls; 
then, at line \autoref{test}, the tested behavior is actually triggered; 
the last part of the test case at \autoref{assertion-1} and \autoref{assertion-2}, the assertion, specifies and checks the conformance of the observed behavior with the expected one.
We note that this notion of call sequence and complex objects is different from test inputs consisting only of primitive values.

% ---------------------------------------------------------------------------------------
% input space exploration: INPUT GOAL
% ---------------------------------------------------------------------------------------
\subsection{Algorithms}

\subsubsection{Input Space Exploration Algorithm}
\label{subsec:input-space-exploration}

\dspot aims at exploring the input space so as to set the program in new, never explored states. To do so, \dspot applies code transformations to the original manually-written test methods. 

\textbf{I-Amplification:} \Iampl, for Input Amplification, is the process of automatically creating new test input points from existing test input points.

\dspot uses three kinds of \Iampl.

\emph{1) Amplification of literals}: the new input point is obtained by changing a literal used in the test (numeric, boolean, string).
For numeric values, there are five operators: $+1$, $-1$, $\times 2$, $ \div 2$, and replacement by an existing literal of the same type, if such literal exists.
For Strings, there are four operators: add a random char, remove a random char, replace a random char and replace the string by a fully random string of the same size.
For booleans, there is only one operator: negate the value;

\emph{2) Amplification of method calls}: \dspot manipulates method calls as follows:
\dspot duplicates an existing method call; removes a method call;
or adds a new invocation for an accessible method with an existing variable as target.

\emph{3) Test objects}:
if a new object is needed as a parameter while amplifying method calls, \dspot creates a new object of the required type using the default constructor if it exists.
In the same way, when a new method call needs primitive value parameters, \dspot generates a random value.

\dspot combines the different kinds of \Iampl iteratively: at each iteration all kinds of \Iampl are applied, resulting in new tests. 
From one iteration to another, \dspot reuses the previously amplified tests, and further applies \Iampl{}.

For example, if we apply an \Iampl on the example presented in \autoref{lst:archetype}, it may generate a new method call on \emph{tl}.
In \autoref{lst:iampl-example}, the added method call is ``removeAll''. Since \dspot changes the state of the program, existing assertions may fail. That is why it removes also all existing assertions.

\begin{lstlisting}[caption={An example of an \Iampl{}: the amplification added a method call to \emph{removeAll()} on \emph{tl}.},label=lst:iampl-example,float,language=java,numbers=left] 
testIterationOrder() {
  TreeList tl=new TreeList(); (*@\label{input-begin-iampl}@*)
  tl.add(1);
  tl.add(2);
  tl.removeAll();(*@\label{input-end-iampl}@*) // method call added
  
  // removed assertions
}
\end{lstlisting}

% ---------------------------------------------------------------------------------------
% Assertion improvement
% ---------------------------------------------------------------------------------------
\subsubsection{Assertion Improvement Algorithm}
\label{subsec:new-assertions}

To improve existing tests, \dspot adds new assertions as follows.

\textbf{\Aampl:} \Aampl, for Assertion Amplification, is the process of automatically creating new assertions.

In \dspot, assertions are added on objects from the original test case, as follows: 
1) it instruments the  test cases to collect the state of a program after execution (but before the assertions), \ie it creates observation points. The state is defined by all values returned by getter methods.
2) it runs the instrumented test to collect the values,
the result of this execution is a map that gives, for each test case object, the values from all getters.
3) it generates new assertions in place of the observation points, using the collected values as oracle. The collected values are used as expected values in the new assertions.
In addition, when a new test input sets the program in a state that throws an exception,  \dspot produces a test asserting that the program throws a specific exception.

For example, let consider \Aampl{} on the test case of the example above. 

First, in \autoref{lst:aampl-example-1} \dspot instruments the test case to collect values, by addding method calls to the objects involved in the test case.

\begin{lstlisting}[caption={In \Aampl{}, the second step is to instrument and run the test to collect runtime values.},label=lst:aampl-example-1,float,language=java,numbers=left] 
testIterationOrder() {
  TreeList tl=new TreeList(); (*@\label{input-begin-aampl}@*)
  tl.add(1);
  tl.add(2);
  tl.removeAll();(*@\label{input-end-aampl}@*)

  // logging current behavior
  Observations.observe(tl.size()); 
  Observations.observe(tl.isEmpty()); 
}
\end{lstlisting}

Second, the test with the added observation points is executed, and subsequently, \dspot{} generates new assertions based on the collected values. On \autoref{lst:aampl-example-2}, we can see that \dspot has generated two new assertions.

\begin{lstlisting}[caption={In \Aampl{}, the last step is to generate the assertions based on the collected values.},label=lst:aampl-example-2,float,language=java,numbers=left] 
testIterationOrder() {
  TreeList tl=new TreeList(); (*@\label{input-begin-aampl2}@*)
  tl.add(1);
  tl.add(2);
  tl.removeAll();(*@\label{input-end-aampl2}@*)

  // generated assertions
  assertEquals(0, tl.size()); // generated assertions
  assertTrue(tl.isEmpty()); // generated assertions
}
\end{lstlisting}

% ---------------------------------------------------------------------------------------
% algorithm
% ---------------------------------------------------------------------------------------
\subsubsection{Test Improvement Algorithm}
\label{subsec:algo}

\begin{algorithm}[t]
\begin{algorithmic}[1]
\REQUIRE{Program $P$}
\REQUIRE{Test Suite $TS$}
\REQUIRE{Amplifiers $amps$ to generate new test data input}
\REQUIRE{$n$ number of iterations of DSpot's main loop}
\ENSURE{An Amplified Test Suite $ATS$}
\STATE{$ATS \leftarrow \emptyset$}
\FOR{$t$ in $TS$}
    \STATE{$U \leftarrow generateAssertions\left(t\right)$}
    \STATE{$ATS \leftarrow \{ x \in U | \mbox{x improves mutation score} \}$}
    \STATE{$TMP \leftarrow ATS$}
    \FOR{$i = 0$ to $n$}
        \STATE{$V \leftarrow [ ]$}
        \FOR{$amp$ in $amps$}
            \STATE{$V \leftarrow V \cup amp.apply\left(TMP\right)$}
        \ENDFOR
        \STATE{$V \leftarrow generateAssertions\left(V\right)$}
        \STATE{$ATS \leftarrow ATS \cup \{ x \in V | \mbox{x improves mutation score} \}$}
        \STATE{$TMP \leftarrow V$}
    \ENDFOR
\ENDFOR
\RETURN{$ATS$}
\end{algorithmic}
\caption{Main amplification loop of \dspot.}
\label{algo:dspot_main}
\end{algorithm}

\autoref{algo:dspot_main} shows the main loop of \dspot. 
\dspot takes as input a program $P$ and its Test Suite $TS$. \dspot also uses an integer $n$ that defines the number of iterations.
\dspot produces an Amplified Test Suite $ATS$, \ie a better version of the input Test Suite $TS$ according to a specific test criterion such as \ms.
For each test case $t$ in the test suite $TS$ (Line 1), \dspot first tries to add assertions without generating any new test input (Line 3),  method $generateAssertions\left(t\right)$ is explained in \autoref{subsec:new-assertions}.
Note that adding missing assertions is the elementary way to improve existing tests.

\dspot initializes a temporary list of tests $TMP$ and applies $n$ times the following steps (Line 6): 
1) it applies each amplifier $amp$ on each tests of $TMP$ to build $V$ (Line 8-9 see \autoref{subsec:input-space-exploration} \ie \Iampl);
2) it generates assertions on generated tests in $V$ (Line 11 see \autoref{subsec:new-assertions}, \ie \Aampl);
3) it keeps the tests that improve the \ms (Line 12).
4) it assigns $V$ to $TMP$ for the next iteration. This is done because even if some amplified test methods in $V$ have not been selected, it can contain amplified test methods that will eventually be better in subsequent iterations.

% ---------------------------------------------------------------------------------------
% Flaky tests elimination
% ---------------------------------------------------------------------------------------
\subsubsection{Flaky tests elimination}
The input space exploration (see \autoref{subsec:input-space-exploration}) may produce test inputs that results in non-deterministic executions.
This means that, between two independent executions, the state of the program is not the same.
Since \dspot generates assertions where the expected value is a hard coded value from a specific run (see \autoref{subsec:new-assertions}), the generated test case may become flaky: it passes or fails depending on the execution and whether the expected value is obtained or not.

To avoid such flaky tests generated by \dspot, we run $n$ times each new test case resulting from amplification ($n$ = 3 in the default configuration). 
If a test fails at least once, \dspot throws it away. 
We acknowledge that this procedure does not guarantee the absence of flakiness. 
However, it gives incremental confidence: if the user wants more confidence, she can tell \dspot to run the amplified tests more times.

% ---------------------------------------------------------------------------------------
% Heuristic of selection
% ---------------------------------------------------------------------------------------

\subsubsection{Selecting Focused Test Cases}
\label{subsubsec:test:cases:selection:for:pr}

DSpot  sometimes produces many tests, from one initial test.
Due to limited time, the developer needs to focus on the most interesting ones.
To select the test methods that are the most likely to be merged in the code base, we implement the following  heuristic.
First, the amplified test methods are sorted according to the ratio of newly killed mutants and the total number of test modifications.
Then, in case of equality, the methods are further sorted according to the maximum numbers of mutants killed in the same method.

The first criterion means that we value short modifications.
The second criterion means that the amplified test method is focused and tries to specify one specific method inside the code.

If an amplified test method is merged in the code base, we consider that the corresponding method as specified. In that case, we do not take into account other amplified test methods that specify the same method.

Finally, in this ordered list, the developer is recommended the amplified tests that are focused, where focus is defined as where at least 50\% of the newly killed mutants are located in a single method. Our goal is to select amplified tests which intent can be easily grasped by the developer: the new test specifies the method.

% ---------------------------------------------------------------------------------------
% Implementation
% ---------------------------------------------------------------------------------------
\subsection{Implementation}

\dspot is implemented in Java.
It consists of 8800+ logical lines of code (as measured by cloc).
For the sake of open-science, \dspot is made publicly available on Github\footnote{\url{https://github.com/STAMP-project/dspot}}.
\dspot uses Spoon\cite{pawlak:hal-01169705} to analyze and transform the tests of the software application under amplification.

\label{subsub:pitest}
In this paper, we aim at improving the \ms of test classes.  In \dspot, we use \pitest\footnote{We use the latest version released: 1.2.0.\url{https://github.com/hcoles/pitest/releases/tag/1.2.0}} because:
1) it targets Java programs, 
2) it is mature and well-regarded,
3) it has an active community. 

An important feature of \pitest is that if the application code remains unchanged, the generated mutants are always the same.
This property is very interesting for test amplification.
Since \dspot only modifies  test code, this feature allows us to compare the \ms of the original test case against the \ms of the amplified version and even compare the absolute number of mutants killed by both test case variants. 
We will exploit this feature in our evaluation.

By default, \dspot uses all the mutation operators available in \pitest: 
    conditionals boundary mutator;
    increments mutator;
    invert negatives mutator;
    math mutator;
    negate conditionals mutator;
    return values mutator;
    void method calls mutator.

%The detailed explanation can be found on the web site\footnote{\url{http://pitest.org/quickstart/mutators/}} of \pitest.

%%%%%%%%%%%%%%%%%%%%%%%%%%%%%%%%%%%%%%%%%%%%%%%%%%%%%%%
% EVALUATION
%%%%%%%%%%%%%%%%%%%%%%%%%%%%%%%%%%%%%%%%%%%%%%%%%%%%%%%
\section{Experimental  Protocol}
\label{sec:protocol}

Automatic test improvement has been evaluated with respect to evolutionary test inputs \cite{tonella} and new assertions \cite{TaoXie2006}. However:
1) the two topics have never been studied in conjunction
2) they have never been studied on large modern Java programs
3) most importantly, the quality of improved tests has never been assessed by developers.

We set up a novel experimental protocol that addresses those three points.
First, the experiment is based on \dspot, which combines test input exploration and assertion generation.
Second, the experiment is made on 10 active \gh projects.
Third, we have proposed improved tests to developers under the form of pull-requests.

We answer the following research questions:

\newcommand\rqpullrequest{RQ1\xspace}
\newcommand\rqcandidates{RQ2\xspace}
\newcommand\rqeffectiveness{RQ3\xspace}
\newcommand\rqAmplVersusIAmpl{RQ4\xspace}

\noindent\textbf{\rqpullrequest}: Are the improved test cases produced by \dspot relevant for developers? Are the developers ready to permanently accept the improved test cases into the test repository?\\
\textbf{\rqcandidates}: To what extent are improved test methods considered as focused?\\
\textbf{\rqeffectiveness}: To what extent do the improved test classes increase the \ms of the original,  manually-written, test classes?\\
\textbf{\rqAmplVersusIAmpl}: What is the relative contribution of \Iampl{} and \Aampl{} to the effectiveness of automatic test improvement?\\

\subsection{Dataset}
\label{subsec:dataset}

We evaluate \dspot by amplifying test classes of large-scale, notable, open-source projects. We include projects that fulfill the following criteria:  
1) the project must be written in Java; 
2) the project must have a test suite based on \junit;
3) the project must be compiled and tested with Maven;
4) the project must have an active community as defined by the presence of pull requests on \gh, see \autoref{subsubsec:answer-rqpullrequest}. 

\begin{landscape}
\begin{table}[ht]
\caption{Dataset of 10 active Github projects considered on our relevance study (RQ1) and quantitative experiments (RQ2, RQ3).}
\begin{tabular}{l|l|r|r|l}
\label{tab:dataset}
project & 
description &
\# LOC & \# PR &considered test classes\\
\hline
\rowcolor[HTML]{EEEEEE}
javapoet & Java source file generator & 
3150 & 93 &
\begin{tabular}{@{}l@{}}
TypeNameTest$^h$~~NameAllocatorTest$^h$\\FieldSpecTest$^l$~~ParameterSpecTest$^l$
\end{tabular}\\
mybatis-3 & Object-relational mapping framework &
20683 & 288 &
\begin{tabular}{@{}l@{}}
MetaClassTest$^h$~~ParameterExpressionTest$^h$\\WrongNamespacesTest$^l$~~WrongMapperTest$^l$
\end{tabular}\\
\rowcolor[HTML]{EEEEEE}
traccar & Server for GPS tracking devices &
32648 & 373 &
\begin{tabular}{@{}l@{}}
GeolocationProviderTest$^h$~~MiscFormatterTest$^h$\\ObdDecoderTest$^l$~~At2000ProtocolDecoderTest$^l$
\end{tabular}\\
stream-lib & Library for summarizing data in streams &
4767 & 21 &
\begin{tabular}{@{}l@{}}
TestLookup3Hash$^h$~~TestDoublyLinkedList$^h$\\TestICardinality$^l$~~TestMurmurHash$^l$
\end{tabular}\\
\rowcolor[HTML]{EEEEEE}
mustache.java & Web application templating system &
3166 & 11 &
\begin{tabular}{@{}l@{}}
ArraysIndexesTest$^h$~~ClasspathResolverTest$^h$\\ConcurrencyTest$^l$~~AbstractClassTest$^l$
\end{tabular}\\
twilio-java & Library for communicating with Twilio REST API &
54423 & 87 &
\begin{tabular}{@{}l@{}}
RequestTest$^h$~~PrefixedCollapsibleMapTest$^h$\\AllTimeTest$^l$~~DailyTest$^l$
\end{tabular}\\
\rowcolor[HTML]{EEEEEE}
jsoup & HTML parser &
10925 & 72 &
\begin{tabular}{@{}l@{}}
TokenQueueTest$^h$~~CharacterReaderTest$^h$\\AttributeTest$^l$~~AttributesTest$^h$
\end{tabular}\\
protostuff& Data serialization library &
4700 & 35 &
\begin{tabular}{@{}l@{}}
TailDelimiterTest$^h$~~LinkBufferTest$^h$\\CodedDataInputTest$^l$~~CodedInputTest$^h$
\end{tabular}\\
\rowcolor[HTML]{EEEEEE}
logback & Logging framework &
15490 & 104 &
\begin{tabular}{@{}l@{}}
FileNamePatternTest$^h$~~SyslogAppenderBaseTest$^h$\\FileAppenderResilience\_AS\_ROOT\_Test$^l$~~Basic$^l$
\end{tabular}\\
retrofit & HTTP client for Android. & 
2743 & 249 &
\begin{tabular}{@{}l@{}}
RequestBuilderAndroidTest$^h$~~CallAdapterTest$^h$\\ExecutorCallAdapterFactoryTest$^h$~~CallTest$^h$
\end{tabular}\\
\end{tabular}
\end{table}
\end{landscape}

We implement those criteria as a query on top of TravisTorrent \cite{msr17challenge}. We randomly selected 10 projects from the result of the query which produces, the dataset presented in \autoref{tab:dataset}.
This table gives the project name, a short description, the number of pull-requests on \gh (\#PR), and the considered test classes.
For instance, javapoet is a strongly-tested and active project, which implements a Java file generator, it has had 93 pull-requests in 2016.

\subsection{Test Case Selection Process for Test-suite Improvement}
\label{subsec:test_preparation}

For each project, we select 4 test classes to be amplified. Those test classes are chosen as follows.

% Unit Tests
First, we select unit-test classes only,
because our approach focuses on unit test amplification. We use the following heuristic to discriminate unit test cases from others: we keep a test class if it executes less than an arbitrary threshold of N statements, \ie if it covers a small portion of the code. In our experiment, we use $N=1500$ based on our initial pilot experiments.

Among the unit-tests, we select 4 classes as follows. Since we want to analyze the performance of \dspot{} when it is provided with both good and bad tests, we select two groups of classes: one group with strong tests, one other group with low quality tests.
We use the \ms to distinguish between good and bad test classes.
Accordingly, our selection process has five steps: 
1) we compute the original \ms of each class with \pitest (see \autoref{subsub:pitest};
2) we discard test classes that have 100\% \ms, because they can already be considered as perfect tests (this is the case for eleven classes, showing that the considered projects in our dataset are really well-tested projects);
3) we sort the classes by \ms ( see \autoref{subsec:metrics}), in ascending order;
4) we split the set of test classes into two groups: high \ms( $> 50\%$) and low \ms  ($< 50\%$);
5) we randomly select 2 test classes in each group.

This selection results with 40 test classes: 24 in high mutation group score and 16 in low \ms group. The imbalance is due to the fact that there are three projects really well tested for which there are none or a single test class with a low \ms (projects protostuff, jsoup, retrofit). Consequently, those three projects are represented with 3 or 4 well-tested classes (and 1 or 0 poorly-tested class). In \autoref{tab:dataset}, the last column contains the name of the selected test classes. Each test class name is indexed by a ``h'' or a ``l'' which means respectively that the class have a high \ms or a low \ms.

\subsection{Metrics}
\label{subsec:metrics}

We use the following metrics during our experiment.

%\textbf{Number of Executed Mutants} ($\#Exec.Mutants$): is the absolute number of mutants executed by a test class. This metric captures the size of the execution trace of the test class under consideration. It is used to discard large integration tests in order to only concentrate on unit tests.

\textbf{Number of Killed Mutants} ($\#Killed.Mutants$): is the absolute number of mutants killed by a test class. We use it to compare the fault detection power of an original test class and the one of its amplified version.

\textbf{Mutation Score}: is the percentage of killed mutants over the number of executed mutants. Mathematically, it is computed as follow:
$$\frac{\#Killed.Mutants}{\#Exec.Mutants} \time 100$$.

\textbf{Increase Killed}: is the relative increase of the number of killed mutants by an original test class $T$ and the number of killed mutants by its amplified version $T_a$.  It is computed as follows:
$$\frac{\#Killed.Mutants_{T_a} - \#Killed.Mutants_T}{\#Killed.Mutants_T}$$
The goal of \dspot is to improve tests such that the number of killed mutants  increases.

\subsection{Methodology}
\label{sec:methodo}
%Can the amplified test cases be included into the test repository maintained by developers?

Our experimental protocol is designed to 
study to what extent the test improvements
are valuable for the developer.

\begin{itemize}
\item \textbf{\rqpullrequest}
% process
To answer to \rqpullrequest, we create pull-request on notable open-source projects.
We automatically improve 19 test classes of real world applications and propose one test improvement to the main developers of each project under consideration. We propose the improvement as a pull request on \gh. A PR is composed of a title, a short text that describes the purpose of changes and a set of code change (aka a patch).
The main developers review, discuss and decide to merge or not each pull request.
We base the answer on the subjective and expert assessment from projects' developers.
If a developer merges an improvement synthesized by \dspot, it validates the relevance of \dspot.
The more developers accept and merge test improvements produced by \dspot into their test suite, the more the amplification is considered successful.

\item \textbf{\rqcandidates{}}
To answer \rqcandidates{}, we compute the number of suggested improvements, to verify that the developer is not overwhelmed with suggestions.
We compute the number of focused amplified test cases, per the technique described in \autoref{subsubsec:test:cases:selection:for:pr}, for each project in the benchmark. We present and discuss the proportion of focused tests out of all proposed amplified tests.

\item \textbf{\rqeffectiveness}
To answer \rqeffectiveness, we see whether the value that is taken as proxy to the developer value -- the mutation score -- is appropriately improved.
For 40 real-world classes, we first run the mutation testing tool \pitest (see \autoref{subsub:pitest}) on the test class. This gives the \ams for this original class. Then, we amplify the test class under consideration and we compute the new \ams after amplification. 
Finally, we compare and analyze the results.

% A-Ampl I-Ampl
\item \textbf{\rqAmplVersusIAmpl}
To answer \rqAmplVersusIAmpl, we compute the number of \Aampl{} and \Iampl{} amplifications. The former means that the suggested improvement is very short hence easy to be accepted by the developer while the latter means that more time would be required to understand the improvement.
First, we collect three series of metrics: 
1) we compute \ams for the original test class; 
2) we improve the test class under consideration using only \Aampl{} and compute the new \ams after amplification; 
3) we improve the test class under consideration using \Iampl{} as well as \Aampl{} (the standard complete \dspot workflow) and compute the \ams after amplification. Then, we compare the increase of \ms obtained by using \Aampl{} only and \Iampl{} + \Aampl{}.\footnote{Note that the relative contribution of \Iampl{} cannot be evaluated alone, because as soon as we modify the inputs in a test case, it is also necessary to change and improve the oracle (which is the role of \Aampl{}).}
\end{itemize}

Research questions 3 and 4 focus on the mutation score to assess the value of amplified test methods. This experimental design choice is guided by our approach to select ``focused'' test methods, which are likely to be selected by the developers (described in \autoref{subsubsec:test:cases:selection:for:pr}). 
Recall that the number of killed mutants by the amplified test is the key focus indicator. Hence, the more \dspot is able to improve the mutation score, the more likely we are to find good candidates for the developers.%This is why we perform this evaluation that generalize the potential of producing a correct input for the heuristic.

% Time
%\subsubsection{RQ5}
%\todo{write it, to Cemetery?}

%%%%%%%%%%%%%%%%%%%%%%%%%%%%%%%%%%%%%%%%%%%%%%%%%%%%%%%
% RESULTS
%%%%%%%%%%%%%%%%%%%%%%%%%%%%%%%%%%%%%%%%%%%%%%%%%%%%%%%
\section{Experimental Results}
\label{sec:results}

We first discuss how automated test improvements done by \dspot are received by developers of notable open-source projects (\rqpullrequest{}).
Then, \rqcandidates{}, \rqeffectiveness{} and \rqAmplVersusIAmpl{} are based on a large scale quantitative experiments over 40 real-world test classes, whose main results are reported in \autoref{tab:overall-results:high_pms}. % and \autoref{tab:overall-results:low_pms}.
For the sake of open-science, all experimental results are made publicly available online:\\
\url{https://github.com/STAMP-project/dspot-experiments/}.

%%%%%%%%%%%%%%%%%%%%%%%%%%%%%%%%%%%%%%%%%%%%
%%%%%%%%%% RQ : case study
%%%%%%%%%%%%%%%%%%%%%%%%%%%%%%%%%%%%%%%%%%%%
% !TEX root = main.tex

\subsection{Answer to \rqpullrequest}
\label{subsubsec:answer-rqpullrequest}

\textbf{\rqpullrequest: Would developers be ready to permanently accept automatically improved test cases into the test repository?}

\subsubsection{Process}

In this research question, our goal is to propose a new test to the lead developers of the open-source projects under consideration. 
The  improved test is proposed through a ``pull-request'', which is a way to reach developers with patches on collaborative development platforms such as Github.

In practice, short pull requests (\ie with small test modifications) with clear purpose, \ie what for it has been opened, have much more chance of being reviewed, discussed and eventually merged. So we aim at providing improved tests which are easy to review.
As shown in \autoref{subsec:input-space-exploration}, \dspot generates several amplified test cases, and we cannot propose them all to the developers.
To select the new test case to be  proposed as a pull request, we look for an amplified test that kills mutants located in the same method.
From the developer's viewpoint, it means that the intention of the test is clear: it specifies the behavior provided by a given method or block.

{The selection of amplified test methods is done as described in \autoref{subsubsec:test:cases:selection:for:pr}. For each selected  method, we compute and minimize the diff between the original method and the amplified one and then we submit the diff as a pull request.
%The selection of amplified test methods to be given as pull-requests is done as follows: 1) we sort the amplified test methods by the ratio between the number of new killed mutants and the number of \Aampl and \Iampl, in descending order; 2) we take the top; 3) we look at the location of the newly killed mutants; 4) if they are all in the same location it means that there is a clear purpose in the amplified test method to specify this previously untested location and thus we keep it
%5) we compute and minimize the diff between the original method and the amplified one;
%6) we submit the diff as a pull-request.
A second point in the preparation of the pull request relates to the length of the amplified test: once a test method has been selected as a candidate pull request, we make the diff as concise as possible for the review to be fast and easy.

\subsubsection{Overview}

In total, we have created 19 pull requests, as shown in \autoref{tab:res-pr}. In this table, the first column is the name of the project, the second is number of  opened pull requests, \ie the number of amplified test methods proposed to developers. The third column is the number of amplified test methods accepted by the developers and permanently integrated in their test suite. The fourth column is the number of amplified test methods rejected by the developers. The fifth column is the number of pull requests that are still being discussed, \ie nor merged nor closed. (This number might change over time if pull-requests are merged or closed.)

\begin{table}[]
\centering
\caption{Overall result of the opened pull request built from result of DSpot.}
\begin{tabular}{l|cccc}
 project & \# opened & \# merged & \# closed & \begin{tabular}{cc} \# under \\ discussion \end{tabular}\\
 \hline
 javapoet & 4 & 4 & 0 & 0\\
 mybatis-3 & 2 & 2 & 0 & 0\\
 traccar & 2 & 1 & 0 & 1\\
 stream-lib & 1 & 1 & 0 & 0\\
 mustache & 2 & 2 & 0 & 0\\
 twilio & 2 & 1 & 0 & 1\\
 jsoup & 2 & 0 & 1 & 1\\
 prostostuff & 2 & 2 & 0 & 0\\
 logback & 2 & 0 & 0 & 2\\
 retrofit & 0 & 0 & 0 & 0\\
 \hline
 total & 19 & 13 & 1 & 5
\end{tabular}
\label{tab:res-pr}
\end{table}

Overall 13 over 19 have been merged. Only 1 has been rejected by developers. There are 5 under discussion.
In the following, we perform a manual analysis of one pull-request per project.
\autoref{tab:list-urls-prs} contains the URLs of pull requests proposed in this experimentation.

\begin{table}[]
\label{tab:list-urls-prs}
\centering
\caption{List of URLs to the pull-requests created in this experiment.}
\begin{tabular}{l|l}
project & pull request urls \\
\hline
javapoet & 
    \begin{tabular}{l}
        \url{https://github.com/square/javapoet/pull/669}\\
        \url{https://github.com/square/javapoet/pull/668}\\
        \url{https://github.com/square/javapoet/pull/667}\\ 
        \url{https://github.com/square/javapoet/pull/544}
    \end{tabular} \\
\hline
mybatis-3 & 
    \begin{tabular}{l}
        \url{https://github.com/mybatis/mybatis-3/pull/1331}\\
        \url{https://github.com/mybatis/mybatis-3/pull/912}
    \end{tabular} \\
\hline
traccar &
    \begin{tabular}{l}
        \url{https://github.com/traccar/traccar/pull/2897}\\
        \url{https://github.com/traccar/traccar/pull/4012}
    \end{tabular} \\
    \hline
stream-lib &
    \begin{tabular}{l}
        \url{https://github.com/addthis/stream-lib/pull/128}\\
    \end{tabular} \\
    \hline
mustache &
    \begin{tabular}{l}
        \url{https://github.com/spullara/mustache.java/pull/210}\\
        \url{https://github.com/spullara/mustache.java/pull/186}
    \end{tabular} \\
    \hline
twilio &
    \begin{tabular}{l}
        \url{https://github.com/twilio/twilio-java/pull/437}\\
        \url{https://github.com/twilio/twilio-java/pull/334}
    \end{tabular} \\
    \hline
jsoup &
    \begin{tabular}{l}
        \url{https://github.com/jhy/jsoup/pull/1110}\\
        \url{https://github.com/jhy/jsoup/pull/840}
    \end{tabular} \\
    \hline
protostuff &
    \begin{tabular}{l}
        \url{https://github.com/protostuff/protostuff/pull/250}\\
        \url{https://github.com/protostuff/protostuff/pull/212}
    \end{tabular} \\
    \hline
logback &
    \begin{tabular}{l}
        \url{https://github.com/qos-ch/logback/pull/424}\\
        \url{https://github.com/qos-ch/logback/pull/365}
    \end{tabular} \\
\end{tabular}
\end{table}

We now present one case study per project of our dataset.

% -----------------------------------------------------------------------------
% JAVAPOET
% -----------------------------------------------------------------------------
\subsubsection{javapoet}

% javapoet used: innerClassInGenericType_cf45194

%\todo{move to dataset Javapoet is a library for generating java source files. The original test suite kills 77.84\% --- 3874/4977 --- of the mutants generated by \pitest. Javapoet is developed by square, a strong and well-known company in the open-source ecosystem. They develop and maintain at least 50 projects on github in diverse language and technologies.}

%\todo{What date and commit id for amplification.}

We have applied \dspot to amplify \texttt{TypeNameTest}. \dspot  synthesizes a single assertion that kills 3 more mutants, all of them at line 197 of the equals method. 
A manual analysis reveals that this new assertion specifies a contract for the method \texttt{equals()} of objects of type \texttt{TypeName}: the method must return false when the input is null. This contract was not tested.

Consequently, we have proposed to the Javapoet developers the following one liner pull request \footnote{\url{https://github.com/square/javapoet/pull/544}}:
\begin{figure}[H]
    \centering
        \fbox{\includegraphics[width=0.98\textwidth, trim=2cm 14.4cm 10.06cm 14.0cm, clip]{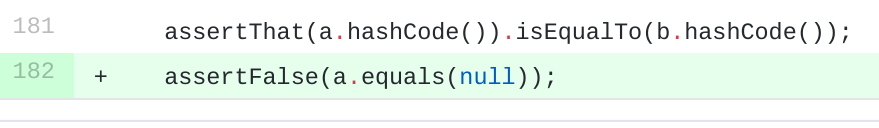}}
\end{figure}
The title of the pull resuest is: ``\emph{Improve test on TypeName}'' with the following short text: ``\emph{Hello, I open this pull request to specify the line 197 in the equals() method of com.squareup.javapoet.TypeName. if (o == null) return false;}''
This test improvement synthesized by DSpot has been merged by of the lead developer of javapoet one hour after its proposal.

% -----------------------------------------------------------------------------
% MYBATIS
% -----------------------------------------------------------------------------
\subsubsection{mybatis-3}

% mybatis-3 used: shouldGetAndSetNestedProperty_cf114673
%\todo{MyBatis is a object relational mapper framework (aka ORM). The original test suite kills 75.3\% --- 14821 over 19685 of the mutant generated by \pitest. DATASET}

In project mybatis-3, We have applied \dspot to amplify a test for \texttt{MetaClass}. \dspot synthesizes a single assertion that kills 8 more mutants.
All new mutants killed are located between lines 174 and 179, \ie the \texttt{then} branch of an \texttt{if-statement} in method \texttt{buildProperty(String property, StringBuilder sb)} of \texttt{MetaClass}.
This method builds a String that represents the  property given as input. The  \texttt{then} branch is responsible to build  the String in case the \texttt{property} has a child, \eg the input is ``richType.richProperty''. This behavior is not specified at all in the original test class.

We have proposed to the developers the following pull request entitled ``\emph{Improve test on MetaClass}'' with the following short text: ``\emph{Hello, I open this pull request to specify the lines 174-179 in the buildProperty(String, StringBuilder) method of MetaClass.}'' \footnote{\url{https://github.com/mybatis/mybatis-3/pull/912/files}}:
\begin{figure}[H]
    \centering\fbox{\includegraphics[width=0.98\textwidth, trim=2cm 17.5cm 4.5cm 10.5cm, clip]{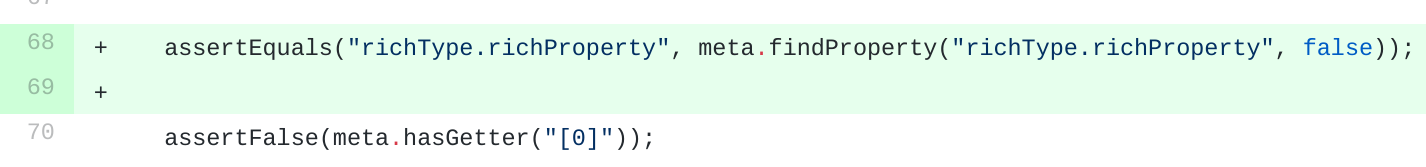}}
\end{figure}

The developer accepted the test improvement and merged the pull request the same day without a single objection.

% -----------------------------------------------------------------------------
% TRACCAR
% -----------------------------------------------------------------------------
\subsubsection{traccar}
% top killer : testDecode_cf39

We have applied \dspot to amplify \texttt{ObdDecoderTest}. It identifies a single assertion that kills 14 more mutants.
All newly killed mutants are located between lines 60 to 80, \ie in the method \texttt{decodesCodes()} of \texttt{ObdDecoder}, which is responsible to decode a \texttt{String}. In this case, the pull request consists of a new test method because the new assertions do not fit with the intent of existing tests. 
This new test method is proposed into \texttt{ObdDecoderTest}, which is the class under amplification. The PR was entitled ``\emph{Improve test cases on ObdDecoder}'' with the following description: ``\emph{Hello, I open this pull request to specify the method decodeCodes of the ObdDecoder}''. \footnote{\url{https://github.com/tananaev/traccar/pull/2897}}
\begin{figure}[H]
    \centering\fbox{\includegraphics[width=0.98\textwidth, trim=2cm 15.57cm 6.35cm 9.88cm, clip]{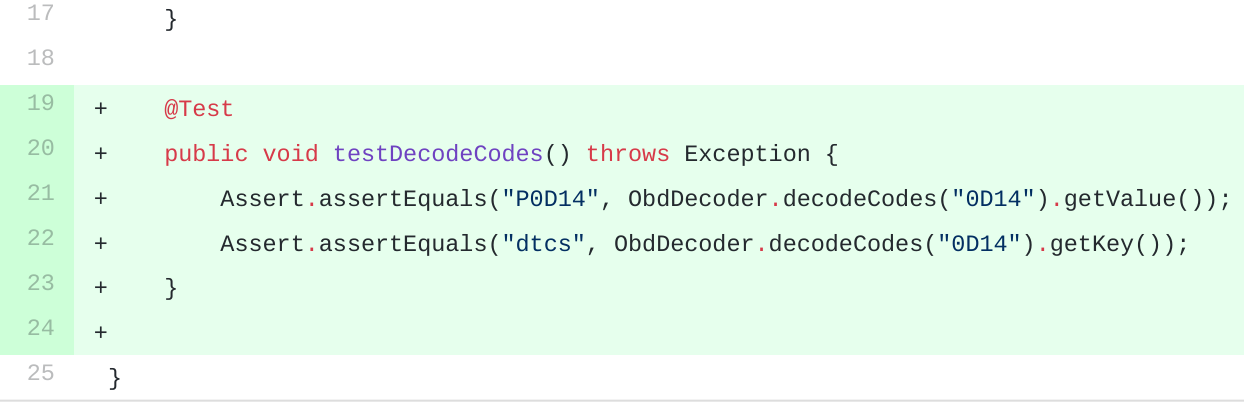}}
\end{figure}

The developer of traccar thanked us for the proposed changes and merged it the same day.

% -----------------------------------------------------------------------------
% STREAM LIB
% -----------------------------------------------------------------------------
\subsubsection{stream-lib}

%testHash64ByteArrayOverload_cf92

% https://github.com/addthis/stream-lib/pull/128/files

We have applied \dspot to amplify \texttt{TestMurmurHash}. It identifies a new test input that kills 15 more mutants.
All newly killed mutants are located in method \texttt{hash64()} of \texttt{MurmurHash} from lines 158 to 216. This method computes a hash for a given array of byte. The PR was entitled ``\emph{Test: Specify hash64}'' with the following description: ``\emph{The proposed change specifies what the good hash code must be. With the current test, any change in "hash" would still make the test pass, incl. the changes that would result in an inefficient hash.}''. \footnote{\url{https://github.com/addthis/stream-lib/pull/127/files}}:
\begin{figure}[H]
    \centering\fbox{\includegraphics[width=0.98\textwidth, trim=2cm 17.07cm 4.90cm 10.59cm, clip]{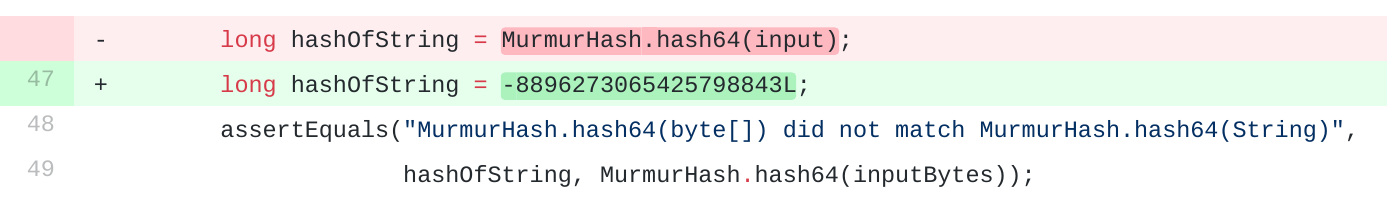}}
\end{figure}

Two days later, one developer mentioned the fact that the test is verifying the overload of the method and is not specifying the method hash itself. He closed the PR because it was not relevant to put changes there. He suggested to open an new pull request with a new test method instead of changing the existing test method. We proposed, 6 days later, a second pull request entitled ``\emph{add test for hash() and hash64() against hard coded values}'' with no description, since we estimated that the developer was aware of our intention.\footnote{\url{https://github.com/addthis/stream-lib/pull/128/files}}:
\begin{figure}[H]
    \centering\fbox{\includegraphics[width=0.98\textwidth, trim=2cm 9.55cm 3.35cm 11.02cm, clip]{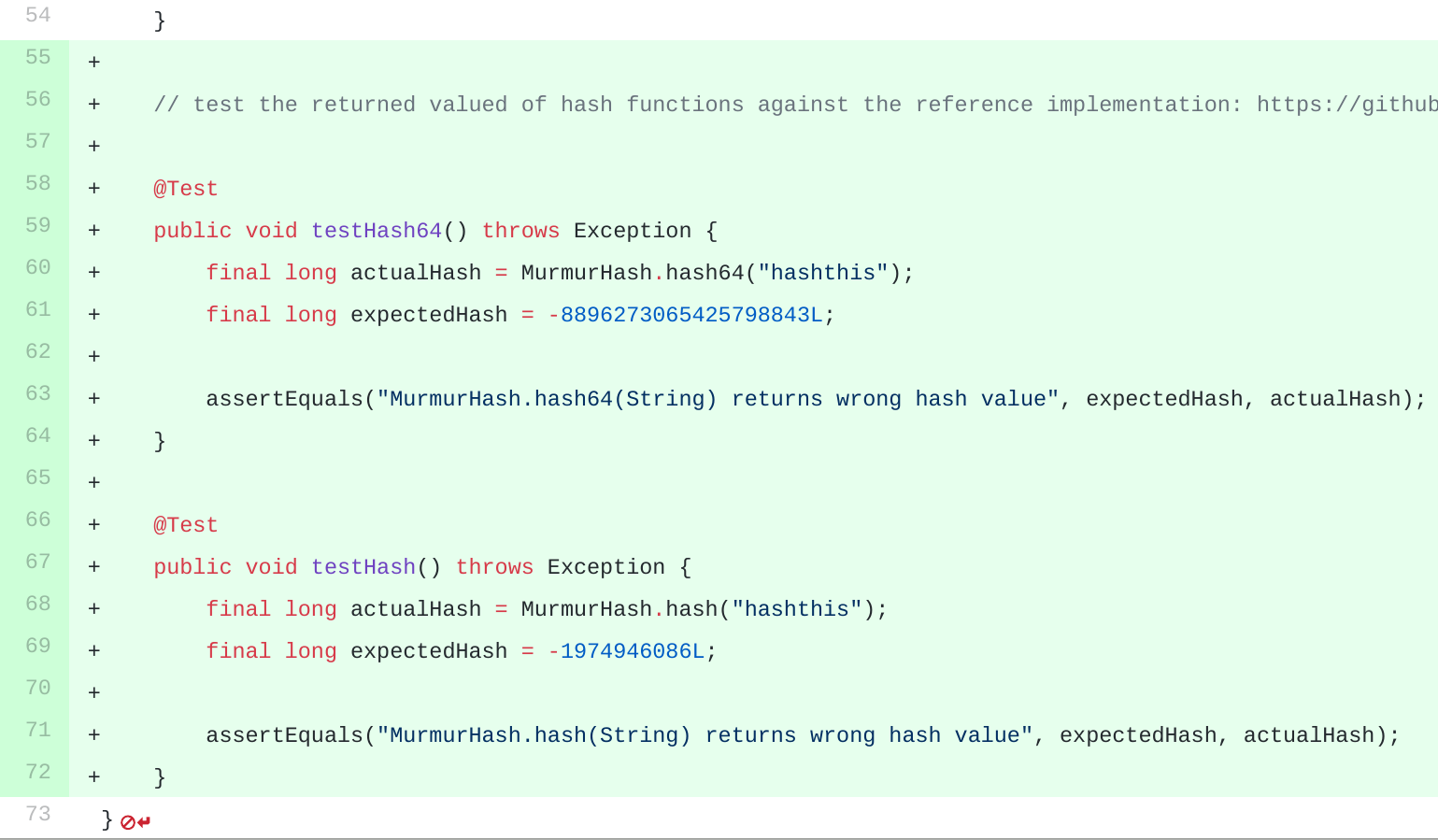}}
\end{figure}

The pull request has been merged by the same developer 20 days later.

% -----------------------------------------------------------------------------
% MUSTACHE.JAVA
% -----------------------------------------------------------------------------
\subsubsection{mustache.java}

%testAbstractClass_add3_literalMutation38_literalMutation326_failAssert19

We have applied \dspot to amplify \texttt{AbstractClassTest}. It identifies a try/catch/fail block that kills 2 more mutants.
This is an interesting new case, compared to the ones previously discussed, because it is about the specification of exceptions, \ie of behavior under erroneous inputs.
All newly killed mutants are located in method \texttt{compile()} on line 194. The test specifies that if a variable is improperly closed, the program must throw a \texttt{MustacheException}. In the Mustache template language, an improperly closed variable occurs when an opening brace ``$\{$'' does not have its matching closing brace such as in the input of the proposed changes. We propose the pull request to the developers, entitled ``\emph{Add Test: improperly closed variable}'' with the following description: ``\emph{Hello, I proposed this change to improve the test on MustacheParser. When a variable is improperly closed, a MustacheException is thrown.}''.\footnote{\url{https://github.com/spullara/mustache.java/pull/186/files}}
\begin{figure}[H]
    \centering\fbox{\includegraphics[width=0.98\textwidth, trim=2cm 8.86cm 2.03cm 14.96cm, clip]{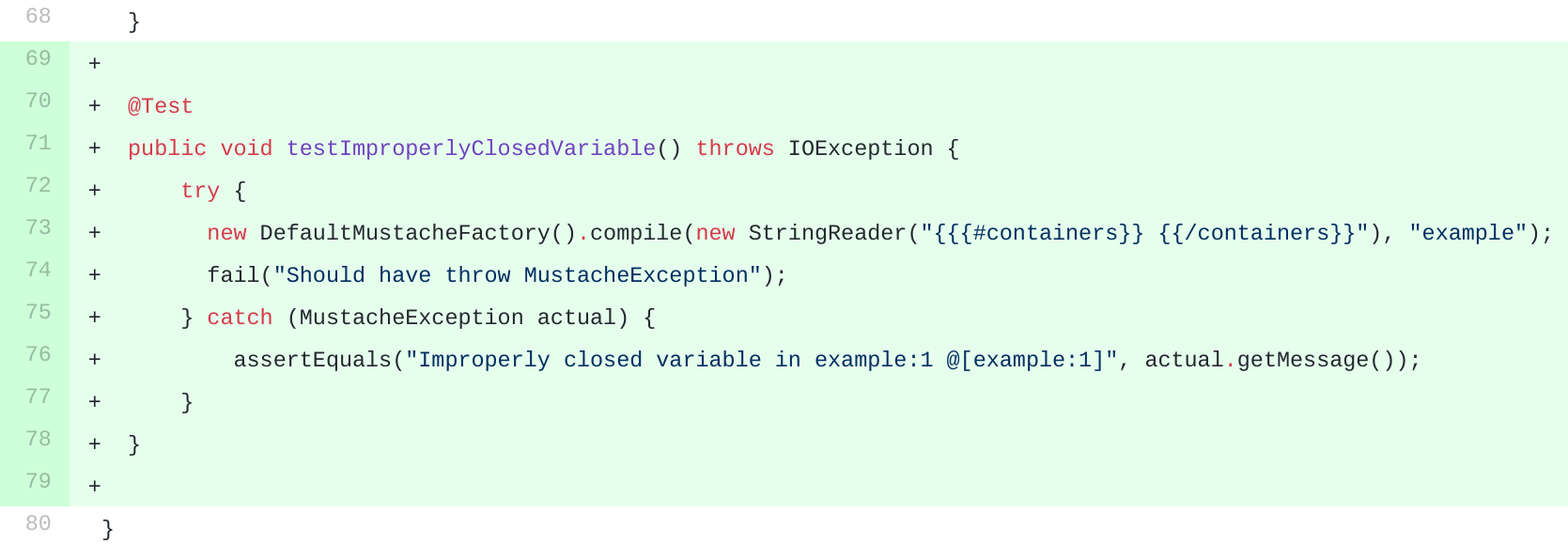}}
\end{figure}

12 days later, a developer accepted the change, but noted that the test should be in another class. He closed the pull request and added the changes himself into the desired class.\footnote{the diff is same:\url{https://github.com/spullara/mustache.java/commit/9efa19d595f893527ff218683e70db2ae4d8fb2d}}. 

% -----------------------------------------------------------------------------
% TWILIO-JAVA
% -----------------------------------------------------------------------------
\subsubsection{twilio-java}

We have applied \dspot to amplify \texttt{RequestTest}. It identifies two new assertions that kill 4 more mutants. All mutants were created between lines 260 and 265 in the method \texttt{equals()} of \texttt{Request}. The change specifies that an object \texttt{Request} is not equal to null nor an object of different type, \ie \texttt{Object} here. The pull request was entitled ``\emph{add test equals() on request}'', accompanied with the short description ``\emph{Hi, I propose this change to specify the equals() method of com.twilio.http.Request, against object and null value}'' \footnote{\url{https://github.com/twilio/twilio-java/pull/334/files}}:
\begin{figure}[H]
    \centering\fbox{\includegraphics[width=0.98\textwidth, trim=2cm 14.83cm 8.11cm 10.11cm, clip]{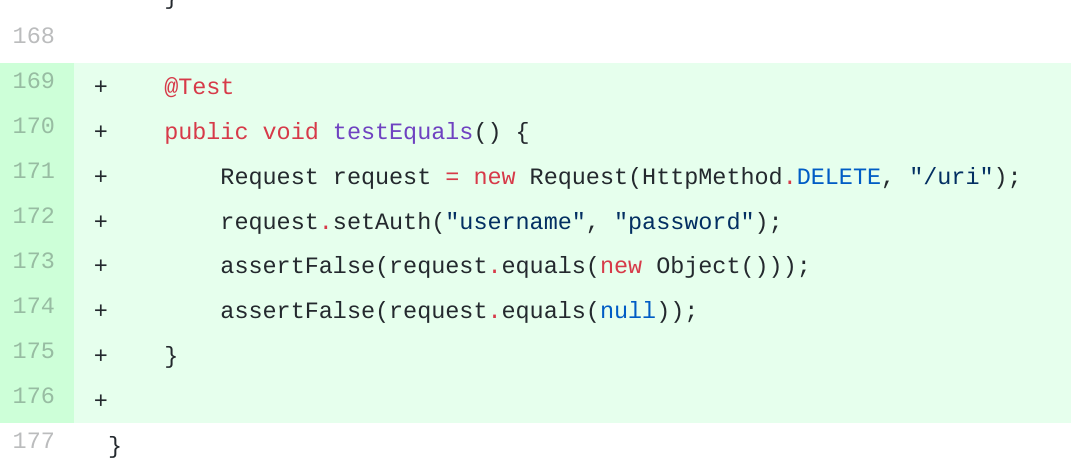}}
\end{figure}

A developer merged the change 4 days later.

% 
% 25 minutes
% testGetUsername_cf47548_cf49619_cf53871
% best fitness, all mutant are killed in equals pr easy to build
% adding assertion manually against null, for having a complete test case.

% ExcludedClasses = com.twilio.http.NetworkHttpClientTest

% -----------------------------------------------------------------------------
% JSOUP
% -----------------------------------------------------------------------------
\subsubsection{jsoup}

We have applied \dspot to amplify \texttt{AttributeTest}. It identifies one assertion that kills 13 more mutants. All mutants are in the method \texttt{hashcode} of Attribute. The pull request was entitled ``\emph{add test case for hashcode in attribute}'' with the following short description ``\emph{Hello, I propose this change to specify the hashCode of the object org.jsoup.nodes.Attribute.}''\footnote{\url{https://github.com/jhy/jsoup/pull/840}}:
\begin{figure}[H]
    \centering\fbox{\includegraphics[width=0.98\textwidth, trim=2cm 15.5cm 9cm 10.4cm, clip]{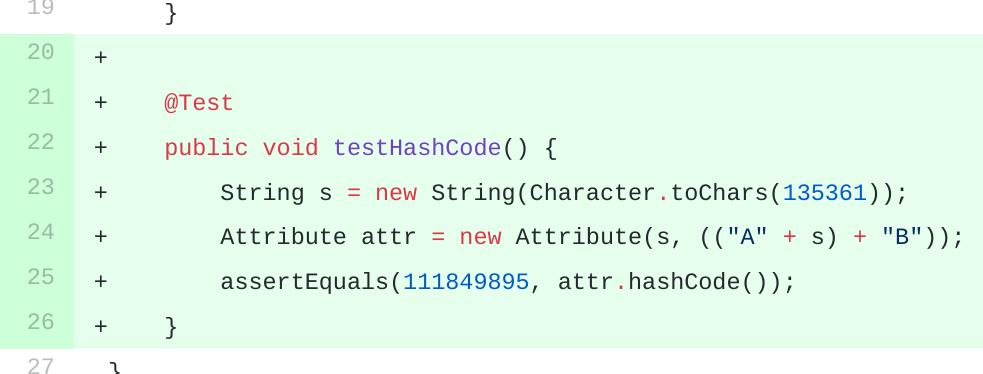}}
\end{figure}

One developer highlighted the point that the \texttt{hashCode} method is an implementation detail, and it is not a relevant element of the API. Consequently, he did not accept our test improvement.

At this point, we have made two pull requests targeting \texttt{hashCode} methods. One accepted and one rejected. \texttt{hashCode} methods could require a different testing approach to validate the number of potential collisions in a collection of objects rather than checking or comparing the values of a few objects created for one explicit test case. The different responses we obtained reflect the fact that developer teams and policies ultimately decide how to test the hash code protocol and the outcome could be different from different projects.

%\TODO{I don't know if i disagree with this developer now, because the contract should be tested not the values.}
%We disagree with this developer, and think that \texttt{hashCode} methods are worth a test case, because the contracts of \texttt{hashCode} methods are essential for many high-level usages such as collection usages. However, we would agree that such generic test cases may be generated by a dedicated testing framework.

% -----------------------------------------------------------------------------
% PROTOSTUFF
% -----------------------------------------------------------------------------
\subsubsection{protostuff}

We have applied \dspot to amplify \texttt{TailDelimiterTest}. It identifies a single assertion that kills 3 more mutants.
All new mutants killed are in the method \texttt{writeTo} of \texttt{ProtostuffIOUtil} on lines 285 and 286, which is responsible to write a buffer into a given scheme. We proposed a pull request entitled ``\emph{assert the returned value of writeList}'', with the following short description ``\emph{Hi, I propose the following changes to specify the line 285-286 of io.protostuff.ProtostuffIOUtil.}''\footnote{\url{https://github.com/protostuff/protostuff/pull/212/files}}, shown earlier in \autoref{fig:diff-protostuff}

\begin{figure}[H]
    \centering\fbox{\includegraphics[width=0.98\textwidth, trim=2cm 16.3cm 4cm 10.4cm, clip]{protostuff.pdf}}
\end{figure}

A developer accepted the proposed changes the same day.

%testEmptyList_add5651

% -----------------------------------------------------------------------------
% LOGBACK
% -----------------------------------------------------------------------------
\subsubsection{logback}

We have applied \dspot to amplify \texttt{FileNamePattern}. It identifies a single assertion that kills 5 more mutant. Newly killed mutants were located at lines 94, 96 and 97 of the \texttt{equals} method of the \texttt{FileNamePattern} class. The proposed pull request was entitle ``\emph{test: add test on equals of FileNamePattern against null value}'' with the following short description: ``\emph{Hello, I propose this change to specify the equals() method ofFileNamePattern against null value}''.\footnote{\url{https://github.com/qos-ch/logback/pull/365/files}}:

\begin{figure}[H]
    \centering\fbox{\includegraphics[width=0.98\textwidth, trim=2cm 11.66cm 8.69cm 14.20cm, clip]{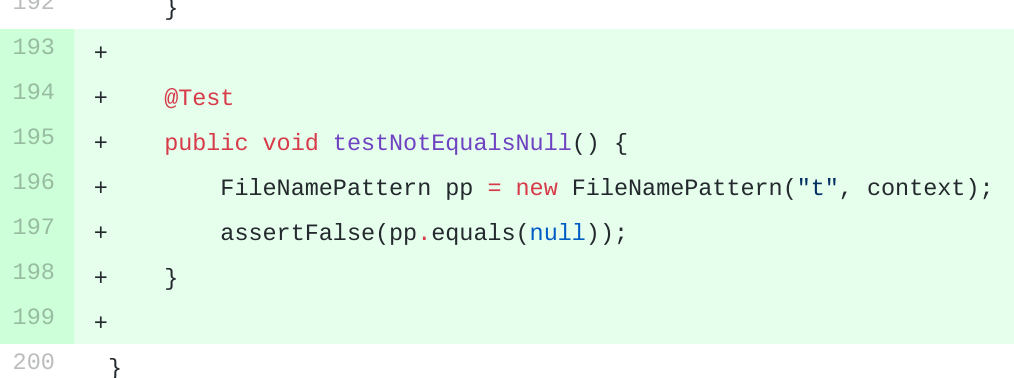}}
\end{figure}

Even if the test asserts the contract that the \texttt{FileNamePattern} is not equals to null, and kills 5 more mutants, the lead developer does not get the point to test this behavior. The pull request has not been accepted.

%testSmoke_cf69003

% -----------------------------------------------------------------------------
% RETROFIT
% -----------------------------------------------------------------------------
\subsubsection{retrofit}

We did not manage to create a pull request based on the amplification of the test suite of retrofit. According to the result, the newly killed mutants are spread over all the code, and thus the amplified methods did not identify a missing contract specification. This could be explained by two facts: 1) the original test suite of retrofit is strong: there is no test class with low \ms and a lot of them are very high \ms, \ie 90\% and more; 2) the original test suite of retrofit uses complex test mechanism such as mock and fluent assertions of the form the \texttt{assertThat().isSomething()}. For the former point, it means that \dspot has been able to improve, even a bit, the \ms of a very strong test suite, but not in targeted way that makes sense in a pull request. For the latter point, this puts in evidence the technical challenge of amplifying fluent assertions and mocking mechanisms.

~\\

%% new section for rev2.13
\subsubsection{Contributions of \Aampl and \Iampl to the Pull-requests}

\begin{table}[]
\caption{Contributions of \Aampl and \Iampl on the amplified test method used to create a pull request.}
\label{tab:contrib-a-i-ampl}
\centering\begin{tabular}{lcc}
\hline
Project & \#\Aampl &  \#\Iampl \\
\hline
javapoet & 2 & 2 \\
mybatis-3 & 3 & 3 \\
traccar & 10 & 7 \\
stream-lib & 2 & 2 \\
mustache & 4 & 3 \\
twilio & 3 & 4 \\
jsoup & 34 & 0 \\
protostuff & 1 & 1 \\
logback & 2 & 2 \\
\hline
\end{tabular}
\end{table}

In \autoref{tab:contrib-a-i-ampl}, we summarize the contribution of \Aampl and \Iampl, where a contribution means an source code modification added during the main amplification loop. In 8 cases over the 9 pull-requests, both \Aampl and \Iampl were necessary. Only the pull request on jsoup was found using only \Aampl. This means that for all the other pull-requests, the new inputs were required to be able: 1) to kill new mutants and 2) to obtain amplified test methods that have values for the developers.

Note that this does not contradict with the fact that the pull-requests are one-liners. Most one-liner pull-requests contain both a new assertion and a new input. Consider the following Javapoet's one liner \texttt{assertFalse(x.equals(null))} (javapoet). In this example, although there is a single line starting with ``assert'', there is indeed a new input, the value ``null''.

~\\
~\\
\begin{mdframed}
\textit{\rqpullrequest: Would developers be ready to permanently accept improved test cases into the test repository?}\\
Answer: We have proposed 19 test improvements to developers of notable open-source projects. 13/19 have been considered valuable and have been merged into the main test suite. The developers' feedback has confirmed the relevance, and also the challenges of automated test improvement.
\end{mdframed}

In the area of automatic test improvement, this experiment is the first to put real developers in the loop, by asking them about the quality of automatically improved test cases. To our knowledge, this is the first public report of automatically improved tests accepted by unbiased developers and merged in the master branch of open-source repositories.

\begin{table}
\caption{The effectiveness of test amplification with \dspot on 40 test classes: 24 well-tested (upper part) and 16 average-tested (lower part) real test classes from notable open-source Java projects.}
%A star after a test name means that the test is considered as subject for the relevance study of \rqpullrequest.}
\label{tab:overall-results:high_pms}
\def\arraystretch{0.55}%  1 is the default, change whatever you need
\setlength\tabcolsep{0.45pt} % default value: 6pt
\small
\begin{tabular}{|llrrrr|rrrr|rrr|r|}
 \rotverticalinv{ID}&
 \rotverticalinv{Class}&
 \rotverticalinv{\# Orig. test methods}&
 \rotverticalinv{Mutation Score}&
 \rotverticalinv{\# New test methods}&
 \rotverticalinv{\begin{tabular}{l}
 Candidates \\ for pull request
 \end{tabular}}&
 \rotverticalinv{\# Killed mutants orig.}&
 \rotverticalinv{\# Killed mutants ampl.}&
 \rotverticalinv{Increase killed}&
 &%arrow for killed
 \setlength{\tabcolsep}{0cm} 
 \rotverticalinv{\begin{tabular}{l}
 \# Killed mutants only\\ A-ampl
 \end{tabular}}&
 \setlength{\tabcolsep}{0cm} 
 \rotverticalinv{\begin{tabular}{l}
 Increase killed only\\ A-ampl
 \end{tabular}}&
 &%arrow for killed
 \rotverticalinv{Time (minutes)}\\
\hline\\
    &\multicolumn{3}{l}{High \ms}\\
\hline\\
1&\scriptsize{TypeNameTest}&12&50\%&19&8&599&715&19\%&{\color{ForestGreen}$\nearrow$}&599&0.0\%&$\rightarrow$&11.11 \\
\rowcolor[HTML]{EFEFEF}
2&\scriptsize{NameAllocatorTest}&11&87\%&0&0&79&79&0.0\%&$\rightarrow$&79&0.0\%&$\rightarrow$&4.76 \\
3&\scriptsize{MetaClassTest}&7&58\%&108&10&455&534&17\%&{\color{ForestGreen}$\nearrow$}&455&0.0\%&$\rightarrow$&235.71 \\
\rowcolor[HTML]{EFEFEF}
4&\scriptsize{ParameterExpressionTest}&14&91\%&2&2&162&164&1\%&{\color{ForestGreen}$\nearrow$}&162&0.0\%&$\rightarrow$&25.93 \\
5&\scriptsize{ObdDecoderTest}&1&80\%&9&2&51&54&5\%&{\color{ForestGreen}$\nearrow$}&51&0.0\%&$\rightarrow$&2.20 \\
\rowcolor[HTML]{EFEFEF}
6&\scriptsize{MiscFormatterTest}&1&72\%&5&5&42&47&11\%&{\color{ForestGreen}$\nearrow$}&42&0.0\%&$\rightarrow$&1.21 \\
7&\scriptsize{TestLookup3Hash}&2&95\%&0&0&464&464&0.0\%&$\rightarrow$&464&0.0\%&$\rightarrow$&6.76 \\
\rowcolor[HTML]{EFEFEF}
8&\scriptsize{TestDoublyLinkedList}&7&92\%&1&1&104&105&0.97\%&{\color{ForestGreen}$\nearrow$}&104&0.0\%&$\rightarrow$&3.03 \\
9&\scriptsize{ArraysIndexesTest}&1&53\%&15&4&576&647&12\%&{\color{ForestGreen}$\nearrow$}&586&1\%&{\color{ForestGreen}$\nearrow$}&10.58 \\
\rowcolor[HTML]{EFEFEF}
10&\scriptsize{ClasspathResolverTest}&10&67\%&0&0&50&50&0.0\%&$\rightarrow$&50&0.0\%&$\rightarrow$&4.18 \\
11&\scriptsize{RequestTest}&17&81\%&4&3&141&156&10\%&{\color{ForestGreen}$\nearrow$}&141&0.0\%&$\rightarrow$&60.55 \\
\rowcolor[HTML]{EFEFEF}
12&\scriptsize{PrefixedCollapsibleMapTest}&4&96\%&0&0&54&54&0.0\%&$\rightarrow$&54&0.0\%&$\rightarrow$&13.28 \\
13&\scriptsize{TokenQueueTest}&6&69\%&18&6&152&165&8\%&{\color{ForestGreen}$\nearrow$}&152&0.0\%&$\rightarrow$&15.61 \\
\rowcolor[HTML]{EFEFEF}
14&\scriptsize{CharacterReaderTest}&19&79\%&71&9&309&336&8\%&{\color{ForestGreen}$\nearrow$}&309&0.0\%&$\rightarrow$&57.06 \\
15&\scriptsize{TailDelimiterTest}&10&71\%&1&1&381&384&0.79\%&{\color{ForestGreen}$\nearrow$}&381&0.0\%&$\rightarrow$&12.90 \\
\rowcolor[HTML]{EFEFEF}
16&\scriptsize{LinkBufferTest}&3&48\%&12&7&66&90&36\%&{\color{ForestGreen}$\nearrow$}&66&0.0\%&$\rightarrow$&3.24 \\
17&\scriptsize{FileNamePatternTest}&12&58\%&27&9&573&686&19\%&{\color{ForestGreen}$\nearrow$}&573&0.0\%&$\rightarrow$&25.08 \\
\rowcolor[HTML]{EFEFEF}
18&\scriptsize{SyslogAppenderBaseTest}&1&95\%&1&1&143&148&3\%&{\color{ForestGreen}$\nearrow$}&143&0.0\%&$\rightarrow$&7.88 \\
19&\scriptsize{RequestBuilderAndroidTest}&2&99\%&0&0&513&513&0.0\%&$\rightarrow$&513&0.0\%&$\rightarrow$&0.04 \\
\rowcolor[HTML]{EFEFEF}
20&\scriptsize{CallAdapterTest}&4&94\%&0&0&55&55&0.0\%&$\rightarrow$&55&0.0\%&$\rightarrow$&7.30 \\
\hline
&\multicolumn{3}{l}{Low \ms}\\
\hline
21&\scriptsize{FieldSpecTest}&2&31\%&12&4&223&316&41\%&{\color{ForestGreen}$\nearrow$}&223&0.0\%&$\rightarrow$&4.44 \\
\rowcolor[HTML]{EFEFEF}
22&\scriptsize{ParameterSpecTest}&2&32\%&11&5&214&293&36\%&{\color{ForestGreen}$\nearrow$}&214&0.0\%&$\rightarrow$&3.66 \\
23&\scriptsize{WrongNamespacesTest}&2&8\%&6&1&78&249&219\%&{\color{ForestGreen}$\nearrow$}&249&219\%&{\color{ForestGreen}$\nearrow$}&29.70 \\
\rowcolor[HTML]{EFEFEF}
24&\scriptsize{WrongMapperTest}&1&8\%&3&1&97&325&235\%&{\color{ForestGreen}$\nearrow$}&325&235\%&{\color{ForestGreen}$\nearrow$}&7.13 \\
25&\scriptsize{ProgressProtocolDecoderTest}&1&16\%&2&1&18&27&50\%&{\color{ForestGreen}$\nearrow$}&23&27\%&{\color{ForestGreen}$\nearrow$}&1.30 \\
\rowcolor[HTML]{EFEFEF}
26&\scriptsize{IgnitionEventHandlerTest}&1&22\%&0&0&13&13&0.0\%&$\rightarrow$&13&0.0\%&$\rightarrow$&0.77 \\
27&\scriptsize{TestICardinality}&2&7\%&0&0&19&19&0.0\%&$\rightarrow$&19&0.0\%&$\rightarrow$&2.13 \\
\rowcolor[HTML]{EFEFEF}
28&\scriptsize{TestMurmurHash}&2&17\%&40&2&52&275&428\%&{\color{ForestGreen}$\nearrow$}&174&234\%&{\color{ForestGreen}$\nearrow$}&2.18 \\
29&\scriptsize{ConcurrencyTest}&2&28\%&2&0&210&342&62\%&{\color{ForestGreen}$\nearrow$}&210&0.0\%&$\rightarrow$&315.56 \\
\rowcolor[HTML]{EFEFEF}
30&\scriptsize{AbstractClassTest}&2&34\%&28&4&383&475&24\%&{\color{ForestGreen}$\nearrow$}&405&5\%&{\color{ForestGreen}$\nearrow$}&12.67 \\
31&\scriptsize{AllTimeTest}&3&42\%&0&0&163&163&0.0\%&$\rightarrow$&163&0.0\%&$\rightarrow$&0.02 \\
\rowcolor[HTML]{EFEFEF}
32&\scriptsize{DailyTest}&3&42\%&0&0&163&163&0.0\%&$\rightarrow$&163&0.0\%&$\rightarrow$&0.02 \\
33&\scriptsize{AttributeTest}&2&36\%&33&11&178&225&26\%&{\color{ForestGreen}$\nearrow$}&180&1\%&{\color{ForestGreen}$\nearrow$}&10.76 \\
\rowcolor[HTML]{EFEFEF}
34&\scriptsize{AttributesTest}&5&52\%&9&6&316&322&1\%&{\color{ForestGreen}$\nearrow$}&316&0.0\%&$\rightarrow$&6.21 \\
35&\scriptsize{CodedDataInputTest}&1&1\%&0&0&5&5&0.0\%&$\rightarrow$&5&0.0\%&$\rightarrow$&3.58 \\
\rowcolor[HTML]{EFEFEF}
36&\scriptsize{CodedInputTest}&1&27\%&29&28&108&166&53\%&{\color{ForestGreen}$\nearrow$}&108&0.0\%&$\rightarrow$&0.88 \\
37&\scriptsize{FileAppenderResilience\_AS\_ROOT\_Test}&1&4\%&0&0&4&4&0.0\%&$\rightarrow$&4&0.0\%&$\rightarrow$&0.65 \\
\rowcolor[HTML]{EFEFEF}
38&\scriptsize{Basic}&1&10\%&0&0&6&6&0.0\%&$\rightarrow$&6&0.0\%&$\rightarrow$&0.89 \\
39&\scriptsize{ExecutorCallAdapterFactoryTest}&7&62\%&0&0&119&119&0.0\%&$\rightarrow$&119&0.0\%&$\rightarrow$&0.09 \\
\rowcolor[HTML]{EFEFEF}
40&\scriptsize{CallTest}&35&69\%&3&1&642&644&0.32\%&{\color{ForestGreen}$\nearrow$}&642&0.0\%&$\rightarrow$&52.84 \\
\hline
\end{tabular}
\end{table}

%%%%%%%%%%%%%%%%%%%%%%%%%%%%%%%%%%%%%%%%%%%%
%%%%%%%%%% RQ : candidates
%%%%%%%%%%%%%%%%%%%%%%%%%%%%%%%%%%%%%%%%%%%%

\subsection{Answer to \rqcandidates{}}

\textbf{\rqcandidates{} To what extent are improved test methods considered as focused?}

% presentation table
\autoref{tab:overall-results:high_pms}
presents the results for RQ2, RQ3 and RQ4.% and \autoref{tab:overall-results:low_pms}
It is structured as follows.
The first column is a numeric identifier that eases reference from the text.
The second column is the name of test class to be amplified.
The third column is the number of test methods in the original test class.
The fourth column is the \ms of the original test class.
The fifth is the number of test methods generated by \dspot.
The sixth is the number of amplified test methods that met the criteria explained in \autoref{subsubsec:test:cases:selection:for:pr}.
The seventh, eight and ninth are respectively the \ams of the original test class, the \ams of its amplified version and the absolute increase obtained with amplification, which is represented with a pictogram indicating the presence of improvement. 
The tenth and eleventh columns concern the \ams when only A-amplification is used.
The twelfth is the time consumed by \dspot to amplify the considered test class. 
The upper part of the table is dedicated to test classes that have a high \ms and the lower for the test classes that have low \ms.

For \rqcandidates{}, the considered results are in the sixth column of \autoref{tab:overall-results:high_pms}. Our selection technique produces candidates that are focused in 25/26 test classes for which there are improved tests. 
For instance, considering test class TypeNameTest (\#8), there are 19 improved test methods, and among them, 8 are focused per our definition and are worth considering to be integrated in the codebase.
On the contrary, for test class ConcurrencyTest (\#29), the technique cannot find any improved test method that matches the focus criteria presented in \autoref{subsubsec:test:cases:selection:for:pr}. In this case, that improved test methods kill additional mutants in 27 different locations. Consequently, the intent of the new amplified tests can  hardly be considered as clear.

Interestingly, for 4 test classes, even if there are more than one improved test methods, the selection technique only  returns one focus candidate (\#23, \#24, \#25, \#40). 
In those cases, there are two possible different reasons:
1) there are several focused improved tests, yet they all specify the same application method (this is the case for \#40
2) there is only one improved test method that is focused (this is the case for \#23, \#24, and \#25)

To conclude, according to this benchmark, \dspot{} proposes at least one and focused improved test in all but one cases. From the developer viewpoint, \dspot is not overwhelming it proposes a small set of suggested test changes, which are ordered, so that even with a small time budget to improve the tests, the developer is pointed to the most interesting case.

~\\
\begin{mdframed}
\textit{\rqcandidates{}:  To what extent are improved test methods considered as focused?}\\
Answer: In 25/26 cases, the improvement is successful at producing at least one focused test method, which is important to save valuable developer time in analyzing the suggested test improvements.
\end{mdframed}
~\\

%%%%%%%%%%%%%%%%%%%%%%%%%%%%%%%%%%%%%%%%%%%%
%%%%%%%%%% RQ amplification
%%%%%%%%%%%%%%%%%%%%%%%%%%%%%%%%%%%%%%%%%%%%

%To what extent does an amplified test class kills more mutants than a manual test class?
\subsection{Answer to \rqeffectiveness}

\textbf{\rqeffectiveness: To what extent do improved test classed kill more mutants than developer-written test classes?}

% 17 + 8
In 26 out of 40 cases, \dspot{} is able to amplify existing test cases and improves the \ms ($MS$) of the original test class.
 % example
For example, let us consider the first row, corresponding to \texttt{TypeNameTest}. This test class originally includes 12 test methods that kill 599 mutants. The improved, amplified version of this test class kills 715 mutants, \ie{} 116 new mutants are killed. This corresponds to an increase of 19\% in the number of killed mutants.

% STRONG TEST
We first discuss the amplification of the test classes that can be considered as being already good tests since they originally have a high \ms: those good test classes are the 24 tests in \autoref{tab:overall-results:high_pms}.
There is a positive increase of killed mutants for 17 cases. This means that even when human developers write good test cases, \dspot{} is able to improve the quality of these test cases by increasing the number of mutants killed. 
In addition, in 15 cases, when the amplified tests kill more mutants, this goes along with an increase of the number of expressions covered with respect to the original test class.

For those 24 well-test classes, the increase in killed mutants varies from 0,3\%, up to 53\%. A remarkable aspect of these results is that \dspot{} is able to improve test classes that are initially extremely strong, with an original \ms of 92\% (ID:8) or even 99\% (ID:20 and ID:21). The improvements in these cases clearly come from the double capacity of \dspot{} at exploring more behaviors than the original test classes and at synthesizing new assertions.

Still looking to the upper part of \autoref{tab:overall-results:high_pms} (the well-tested classes), we now focus on the relative increase in killed mutants (column ``Increase killed''). 
The two extreme cases are \texttt{CallTest} (ID:24) with a small increase of 0.3\% and \texttt{CodeInputTest} (ID:18) with an increase of 53\%.
\texttt{CallTest} (ID:24) initially includes 35 test methods that kill 69\% of 920 covered mutants. Here, \dspot{} runs for 53 minutes and succeeds in generating only 3 new test cases that kill 2 more mutants than the original test class, and the increase in \ms is only minimal. The reason is that input amplification does not trigger any new behavior and assertion amplification fails to observe new parts of the program state. 
Meanwhile, \dspot{} succeeds in increasing the number of mutants killed by \texttt{CodeInputTest} (ID:18) by 53\%. Considering that the original test class is very strong, with an initial \ms of 60\%, this is a very good achievement for test amplification. In this case, the \Iampl applied easily finds new behaviors based on the original test code.
It is also important to notice that the amplification and the improvement of the test class goes very fast here (only 52 seconds). 

One can notice 4 cases (IDs:3, 13, 15, 24) where the number of new test cases is greater than the number of newly killed mutants. This happens because \dspot{} amplifies test cases with different operators in parallel. While we keep only test cases that kill new mutants, it happens that the same mutant is newly killed by two different amplified tests generated in parallel threads. In this case, \dspot{} keeps both test cases.

%Interpretation not able to kills
There are 7 cases with high \ms for which \dspot{} does not improve the number of killed mutants. In 5 of these cases, the original \ms is greater than 87\% (IDs: 2, 7, 12, 21, 22), and \dspot{} does not manage to synthesize improved inputs to cover new mutants and eventually kill them. In some cases \dspot{} cannot improve the test class because they rely on an external resource (a jar file), or use utility methods that are not considered as test methods by \dspot and hence are not modified by our tool.

% WEAK TESTS
Now we consider the tests in the lower part of \autoref{tab:overall-results:high_pms}. %\autoref{tab:overall-results:low_pms}.  
Those tests are weaker because they have a lower \ms. 
%They might be a better reflection  of tests written by the majority of developers.
When amplifying  weak test classes,  \dspot{} improves the number of killed mutants in  9 out of 16 cases. On a per test class basis, this does not differ much from the well tested classes. However, there is a major difference when one considers the increase itself: the increases in number of killed mutants range from 24\% to 428\%. Also, we observe a very strong distinction between test classes that are greatly improved and test classes that are not improved at all (9 test classes are much improved, 7 test classes cannot be improved at all, the increase is 0\%). In the former case, we find test classes that provide a good seed for amplification. In the latter case, we have test classes that are designed in a way that prevents amplification because they use external processes, or depend on administration permission, shell commands and external data sources; or extensively use mocks or factories; or simply very small test cases that do not provide a good potential to \dspot to perform effective amplification.

~\\
\begin{mdframed}
\textit{\rqeffectiveness: To what extent do improved  test classes kill more mutants than manual test classes?}\\
Answer: In our novel quantitative experiment on automatic test improvement, \dspot significantly improves the capacity of test classes at killing mutants in 26 out 40 of test classes, even in cases where the original test class is already very strong. 
Automatic test improvement works particularly well for weakly tested classes (lower part of \autoref{tab:overall-results:high_pms}): the \ms of three classes is increased by more than 200\%.
%Increases range from 0.22\% to 492.86\%.
\end{mdframed}
~\\
The most notable point of this experiment is that we have considered tests that are already really strong (\autoref{tab:overall-results:high_pms}), with \ms in average of 78\%, with the surprising case of a test class with 99\% \ms that \dspot is able to improve.

%%%%%%%%%%%%%%%%%%%%%%%%%%%%%%%%%%%%%%%%%%%%
%%%%%%%%%% RQ 4
%%%%%%%%%%%%%%%%%%%%%%%%%%%%%%%%%%%%%%%%%%%%
\subsection{Answer to \rqAmplVersusIAmpl}

\textbf{What is the contribution of \Iampl{} and \Aampl{} to the effectiveness of automated test improvement?}

The relevant results are reported in the tenth and eleventh column of \autoref{tab:overall-results:high_pms}. %and \autoref{tab:overall-results:low_pms}.
They give the \ams and the relative increase of the number of killed mutants when only using \Aampl.

% example
For instance, for \texttt{TypeNameTest} (first row, id \#1), using only \Aampl kills 599 mutants, which is exactly the same number of the original test class. In this case, both the absolute and relative increase are obviously zero.
On the contrary, for \texttt{WrongNamespacesTest} (id \#27), using only \Aampl  is very effective, it enables \dspot to kill 249 mutants, which, compared to the 78 originally killed mutants, represents an improvement of 219\%. 

Now, if we aggregate over all test classes, our results indicate that \Aampl only is able to increase the number of mutants killed in 7 / 40 test classes. Increments range from 0.31\% to 13\%. 
Recall that when \dspot runs both \Iampl{} and \Aampl, it increases the number of mutants killed in 26 / 40 test classes, which shows that it is indeed the combination of \Aampl and \Iampl which is effective.

We note that \Aampl{} performs as well as \Iampl + \Aampl in only 2/40 cases (ID:27 and ID:28). In this case, all the improvement comes from the addition of new assertions, and this improvement is dramatic (relative increase of 219\% and 235\%).

The limited impact of \Aampl alone has several causes. First, many assertions in the original test cases are already good and precisely specify the expected behavior for the test case.
Second, it might be due to the limited observability of the program under test (\ie, there is a limited number of points where assertions over the program state can be expressed).
Third, it happens when one test case covers global properties across many methods: test \#28 \texttt{WrongMapperTest} specifies global properties, but is not well suited to observe fine grained behavior with additional assertions. This latter case is common among the weak test classes of the lower part of \autoref{tab:overall-results:high_pms}.
%\autoref{tab:overall-results:low_pms}.

~\\

\begin{mdframed}
\textit{\rqAmplVersusIAmpl: What is the contribution of \Iampl{} and \\\Aampl{} to the effectiveness of test amplification?}\\
Answer: The conjunct run of \Iampl{} and \Aampl{} is the best strategy for \dspot{} to improve manually-written test classes. This experiment has shown that \Aampl{} is ineffective, in particular on tests that are already strong.
\end{mdframed}
~\\

To the best of our knowledge, this experiment is the first to evaluate the relative contribution of \Iampl and \Aampl to the effectiveness of automatic test improvement.

\section{Threats to Validity}
\label{sec:threats}

\textbf{\rqpullrequest{}}
The major threat to \rqpullrequest{} is that there is a potential bias in the acceptance of the proposed pull requests.
For instance, if we propose pull requests to colleagues, they are more likely to merge them.
However, this is not the case here.
In this evaluation, the pull requests are submitted by the first author, who is unknown to all considered projects. 
The developers who study the \dspot pull requests are independent from our group and social network.
Since the first author is unknown for the pull request reviewer, this is not a specific bias towards acceptance or rejection of the pull request.
% \rev{}{\textbf{Manual Minimization.} Before proposing pull request, we manually removes unnecessary statements, renames elements (test method name, variable name, etc..) and integrated into the existing test suite. One major threats of our result is that this process is not yet automatized. However, we created a first version of an algorithm able to remove unnecessary statements and research on naming and integration of amplified test methods are in our plan to make the approach totally automatized. The naming convention could be based on machine learning to recognize convention used in the existing test suite and we created an heuristic that extract the specific intent of amplified test methods.}

\textbf{\rqcandidates{}}
The technique used to select focused candidates is based on the proportion of mutant killed and the absolute number of modification done by  the amplification. However, it may happen that some improvements that are not focused per our definition would still be considered as valuable by developers. Having such false negative is a potential threat to validity.

\textbf{\rqeffectiveness{}}
A threat to \rqeffectiveness{} relates to external validity: if the considered projects and tests are written by amateurs, our findings would not hold for serious software projects.
However, we only consider real-world applications, maintained by professional and esteemed open-source developers. This means we tried to automatically improve tests that are arguably among the best of the open-source world, aiming at as strong construct validity as possible.

\textbf{\rqAmplVersusIAmpl{}.}
The main threat to \rqAmplVersusIAmpl{} relates to internal validity: since our results are of computational nature, a bug in our implementation or experimental scripts may threaten our findings. We have put all our code publicly-available for other researchers to reproduce our experiment and spot the bugs, if any.

\textbf{Oracle.}
% faulty program, faulty amplified test method
\dspot generates new assertions based on the current behavior of the program. If the program contains a bug, the resulting amplified test methods would enforce this bug. This is an inherent threat, inherited from \cite{Xie2006}, which is unavoidable when no additional oracle is available, but only the current version of the program. 
To that extent, the best usage of \dspot is to improve the test suite of a supposedly almost correct version of the program.

% ---------------------------------------------------------------------------------------
% RELATED WORK
% ---------------------------------------------------------------------------------------
\section{Related Work}
\label{sec:related}

This work on test amplification  contributes to the field of genetic improvement (GI)  \cite{petke2017genetic}. 
The key novelty is to consider a test suite as the object to be improved, while previous GI works improve the application code.
(Yet, they use the test suite as a fitness function while  assessing the degree of improvement.) The work of Arcuri and Yao \cite{arcuri2008novel} and Wilkerson \etal \cite{wilkerson2010coevolutionary} are good examples of such work that use the test suite as fitness, while improving the program for automatic bug fixing. Both work follow a similar approach: evolve the input program into new versions that pass the regression test suite and that also pass the bug revealing test case (that fails on the original program). In this paper, we do not evolve the application code but the test code.

Evosuite is a state of the art tool to generate test cases for Java program \cite{fraser2013whole}. Evosuite and \dspot have different goals. Evosuite generates new tests, while \dspot improves existing developer-written tests. The interaction between developers and synthesized tests is key here: in 2016, an empirical study demonstrated that developers who are asked to add oracles in test cases generated by Evosuite, produce test suites that are not better than manually written test suites at detecting bugs \cite{fraser2015does}. On the contrary, \dspot is designed to improve manually written test suites to detect more bugs, and  the relevance study of \rqpullrequest demonstrates that the outcome of \dspot is considered as valuable by developers in order to improve existing test suites. 

Our work is related to previous work that aim at automatically generating test cases to improve the mutation score of a test suite.
Liu \etal \cite{multiple-mutants} aim at generating small test cases, by targeting a path that covers multiple mutants to create test inputs. They evaluate their approach on five small projects. Fraser and Arcuri \cite{evosuite:emse14_mutation} propose a search-based approach to generate test suites that maximize the mutation score. However their work is different from ours since they generate new test cases from scratch, while \dspot always starts from developper-written tests. 
Baudry \etal \cite{Baudry05a} improve the mutation score of test suites using a bacteriological algorithm. They run experiments on a small dataset and confirm that their approach is able to increase the mutation score of tests. However, the scope of the study is limited to small programs, and they do not consider the synthesis of assertions.

Other works aim at increasing fault detection capacities of test suites.
Zhang \etal \cite{Zhang:2016:IRT:2950290.2950313}, propose the Isomorphic Regression Testing system and its implementation in ISON. It considers two versions of a program P and P'(for instance P' is the updated version of P, on which we want to detect any regression). First, ISON identifies isomorphims, that is to say, code fragments that have the same behavior. Then, they run the test suite on P and P' to identify which of the branches are uncovered in the isomorphic part, and they  collect the output. In order to cover all branches, they compute a branch condition to execute the uncovered code. They compare ISON to Evosuite,  and conclude that Evosuite achieves a better branch coverage, while ISON is able to detect faults that Evosuite does not.

Harder \etal \cite{Harder03} start from an existing test suite. They evaluate the quality of this initial test suite with respect to operational abstractions, i.e., an abstract description of the behavior covered by the test suite. Their work is about selecting new valuable tests, while ours is about synthesizing new valuable tests.

Then, they generate novel test cases and keep only the ones that change the operational abstraction.
The new test cases are generated by mining invariants using Daikon.  They evaluate their approach on 8 C programs, and show that it generates test cases with good fault-detection capabilities.

Milani \etal \cite{milani2014} propose an approach which combines the advantages of manually written tests and automatic test generation. They exploit the knowledge of existing tests and then combine it with the power of automated crawling. It has been shown that the approach can effectively improve the fault detection rate of the original test suite.
Test amplification, as considered in this work, is different, as it aims at enhancing the fault detection power of manually written test suites.

Yoo \etal  \cite{Yoo:2012:TDR:2237756.2237758} propose Test Data Regeneration(TDR), which is a kind of test amplification. They use hill climbing on existing test data (set of input) that meets a test objective (\eg cover all branch of a function). The algorithm is based on \emph{neighborhood} and a \emph{fitness} functions as the classical hill climbing algorithm. The goal is to create new test data inputs, that have the same behavior as the original one (\eg cover same branches). 
The key novelties of our work with respect to the work of Yoo \etal \cite{Yoo:2012:TDR:2237756.2237758} are as follow: they mutate only literals in existing test cases, while \dspot's \Iampl also amplifies method calls and can synthesize new objects when needed, \Aampl{} makes the synthesis of assertions an integral part of our test suite improvement process and we evaluate the relevance of the synthesized test cases by proposing them to the developers.

Xie \cite{Xie2006} proposes a technique to add assertions into existing test methods. His approach is similar to what we propose with \Aampl{}. However, this work does not consider the synthesis of new test inputs (\Iampl) and hence cannot cover new execution paths. This is the novelty of \dspot and our experiments showed that this is an essential mechanism to improve the test suite.

We now discuss a group of papers together. Pezz\`e \etal \cite{Pezze:2013:GEI:2510665.2511580} synthesize integration test cases from unit test cases. The idea is to combine unit test cases, which test simple functionalities on specific objects, to create new integration test cases supported by the fact that unit test cases are early developed, and integration test cases require more effort to do so.
R{\"o}$\beta$ler et al. \cite{robetaler2012isolating} aim to isolate failure causes. They propose BugEx, a system that starts from a single failing test as input and generates test cases. It extracts the differences in path execution between failing and passing tests. They evaluate BugEx on 7 failures and show that it is able to lead to the failure root causes in 6 cases.
Yu \etal \cite{Yu2013} augment test suites to enhance fault localization. They use test input transformations to generate new test cases in existing test suites. They transform iteratively some existing failing tests to derive new test cases potentially useful to localize the specific encountered fault, similarly at \Iampl. Their tool is designed to target GUI applications.
To reproduce a crash occurred in production, Xuan \etal \cite{Xuan:2015:CRV:2786805.2803206} propose to transform existing test cases. The approach first selects relevant test cases based on the stack trace in the crash, followed by the elimination of assertions in selected test cases, and finally uses a set of predefined transformations to produce new test cases that can help to reproduce the crash.
None of those works have evaluated whether the technique scales on object-oriented applications of the size considered here, and whether the synthesized tests are considered valuable by senior developers.

It can be noted that several test generation techniques start from a seed and evolve it to produce a good test suite. This is the case for techniques such as concolic test generation \cite{godefroid2005dart}, search-based test generation \cite{fraser2012seed}, or random  test generation \cite{groce2007randomized}. The main difference between all these works and \dspot lies in the nature of the seed: previous work use input values in the form of numerical or String values, vectors or files, and do not consider any form of oracle. On the contrary, we consider as a seed a real test case.
It means the seed is a complete program, which creates objects, manipulates the state of these objects, calls methods on these objects and asserts properties on their behavior. This is the contribution of \dspot: using real and complex object-oriented tests as seed.

Almasi \etal \cite{IndustrialEvalAlmasi2017} investigate the efficiency and effectiveness of automated test generation on a production ready application named \emph{LifeCalc}. They use 25 real faults from \emph{LifeCalc} to evaluate two state-of-the-art tools, Evosuite and Randoop, by asking feedback from the developers about the generated test methods. The result are as follows:  overall the tools found 19 over 25 real faults; 
The developers state that the assertions and the readability of generated test methods must be improved. The developers also suggest that such tools should be implemented in continuous integration.
The reason of the 7 faults that remain undetected is that they either require complex test data input or specific assertions. Our experiment is larger in scope, we evaluate \dspot on 10 notable open-source software from GitHub, by proposing amplified test methods in pull requests.

Allamanis \etal \cite{pull:request:evaluation} devised a technique to rename elements in code and evaluate their approach through five pull requests where four of them have been accepted. Their work and ours both rely on independent evaluation through pull-requests. One important difference is that, in the description of the pull request, they say that the improvements are generated by a tool, while in our case, we did not say anything about the research project underlying our pull requests.

% ---------------------------------------------------------------------------------------
% CONCLUSION
% ---------------------------------------------------------------------------------------
\section{Conclusion}
\label{sec:conclusion}

We have presented \dspot, a novel approach to automatically improve existing developer-written test classes.
We have shown that \dspot is able to strengthen real unit test classes in Java from 10 real-world projects. 
Our experiment with real developers indicates that they are ready to merge test cases improved by \dspot into their test suite. 
The road ahead for automatic synthesis of test case improvements is exciting.

First, there is a need to study how to generate meaningful natural language explanations of the suggested test improvements: generation of well named tests, generation of text accompanying the pull request, we dream of using natural-language deep-learning for this task.

Second, we aim at automating even more the process of integrating the amplification result in a ready-to-use pull request. This requires two major steps: first, one needs to identify which parts of the amplified test methods are ``valuable''. 
Second, we need to choose between modifying an existing test method or create a new one that is derived from an existing one, even if the new method is by construction an extension of an existing one. 
Such a decision procedure must be made based on the intention of the existing test methods and the potentially new intention of the amplified test. If we find an existing test method that carries the same intention, \ie it tests the same portion of code as the amplification, one would preferably add changes there rather than creating a new test methods. This challenging vision of mining and comparing test purposes is the main area of our future work.

Third, and finally, we envision to integrate \dspot in a continuous integration service (CI) where test classes would be amplified on-the-fly for each commit. This would greatly improve the direct industrial applicability of this software engineering research.

\balance
\bibliographystyle{abbrv}
\bibliography{ref}

\end{document}